\documentclass[aps,twocolumn,prb,floatfix,superscriptaddress,amsmath,amssymb]{revtex4-2} %% for PRB
\usepackage{dcolumn}
\usepackage{amsmath}
\usepackage{mathrsfs}
\usepackage{txfonts}
\usepackage{bm}
\usepackage[T1]{fontenc}
\usepackage{xspace}
\usepackage{comment}
\usepackage{braket}
\usepackage{bbold}
\usepackage{mathtools}
\newcommand{\normord}[1]{\mathopen{:}#1\mathclose{:}}
\usepackage{graphicx}
\usepackage{hyperref}
\usepackage{color}
\usepackage{xcolor}
\usepackage[version=3]{mhchem}
\hypersetup{
        colorlinks=true,
        citecolor=blue,
        urlcolor=blue,
        linkcolor=blue
}

\begin{document}
\let\emph\textit

\title{
    Chiral bosonic mean-field Ansatz and spin dynamics in spin-1 Kitaev magnets
}
\author{Daiki Sasamoto}
\email[sasamoto.daiki.r6@dc.tohoku.ac.jp]{}
\affiliation{
  Department of Physics, Graduate School of Science, Tohoku University, Sendai, Miyagi 980-8578, Japan
}
\author{Arnaud Ralko}
\affiliation{Institut N\'eel, Universit\'e Grenoble Alpes and CNRS, Grenoble 38042, France}
\author{Jaime Merino}
\affiliation{Departamento de F\'isica Te\'orica de la Materia Condensada, Condensed Matter Physics Center (IFIMAC) and Instituto Nicol\'as Cabrera, Universidad Aut\'onoma de Madrid, Madrid 28049, Spain}
\author{Joji Nasu}
\affiliation{
  Department of Physics, Graduate School of Science, Tohoku University, Sendai, Miyagi 980-8578, Japan
}

\date{\today}
\begin{abstract}
The Kitaev model is a paradigmatic system for realizing quantum spin liquids, but its higher-spin extensions are not exactly solvable, and their spin dynamics is less well understood than in the spin-$1/2$ case.
For integer spin, recent theoretical developments have motivated bosonic parton descriptions, and Schwinger-boson mean-field theory provides a useful framework for treating bosonic spinons in bond-dependent Kitaev interactions.
In this work, we reexamine a previously introduced triplet-pairing $\phi_t=\pi/2$ phase pattern for the antiferromagnetic $S=1$ Kitaev model and extend the analysis to weak symmetric off-diagonal exchanges $\Gamma$ and $\Gamma^{\prime}$.
Using a bond-operator formulation of Schwinger-boson mean-field theory, we calculate the dynamical spin structure factor for the triplet $0$-flux and triplet $\pi/2$-flux Ans\"atze with a spin-correlation scheme appropriate for Kitaev interactions.
In the full mean-field calculation for the pure Kitaev limit, the $\pi/2$-flux Ansatz yields a flatter spectrum than the $0$-flux Ansatz, consistent with its weakly dispersive spinon bands.
The corresponding real-space spin correlations show that the $\pi/2$-flux Ansatz suppresses longer-distance correlations more strongly than the $0$-flux Ansatz, yielding a correlation pattern closer to the short-ranged form expected in the Kitaev limit.
This comparison shows that the flatness of $S(\bm{q},\omega)$ is closely tied to short-ranged spin correlations and is therefore an important consistency check, although it is not, by itself, a diagnostic of time-reversal-symmetry breaking.
Three-spin correlations reveal staggered chirality in the mean-field state, which breaks time reversal but preserves combined inversion and time reversal.
At the pure Kitaev point, the $\pi/2$-flux Ansatz also captures the dominant low-energy spectral features of the exact-diagonalization results better than the $0$-flux Ansatz.
We then study weak off-diagonal exchanges along $\Gamma^{\prime}=\Gamma$ near the pure Kitaev limit, taking the same-sign relation from analyses of candidate spin-$1$ Kitaev materials.
For representative weak same-sign perturbations, gapped solutions are obtained within the constrained $\pi/2$-flux manifold.
Taken together, these results support the triplet $\pi/2$-flux chiral bosonic Ansatz as a useful mean-field description of spin dynamics near the antiferromagnetic $S=1$ Kitaev limit with weak off-diagonal exchanges.

\end{abstract}
\maketitle

\section{Introduction}
\label{introduction}

Quantum spin liquids are exotic quantum states of matter in which long-range magnetic order is avoided down to zero temperature by strong quantum fluctuations~\cite{Anderson-1973,Ramirez-1994,Balents-2010,Savary-Balents-2016,Zhou-Kanoda-2017,Knolle-Moessner-2019,Wen-Yu-2019,Broholm-Cava-2020,Clark-Abdeldaim-2021}.
Originally, Anderson proposed the quantum spin-liquid state as a candidate ground state of the antiferromagnetic Heisenberg model on the triangular lattice~\cite{Anderson-1973}.
It is now appreciated not merely as a disordered phase but also as a theoretically rich state that can host fractionalized elementary excitations and, when gapped, can possess a topologically nontrivial structure~\cite{Kalmeyer-Laughlin-1987,Wen-1989,Wen-2002,Messio-Lhuillier-2013}.
In particular, the $S=1/2$ Kitaev model defined on the honeycomb lattice~\cite{Kitaev-2006} has dramatically advanced the study of quantum spin liquids.
Its ground state is rigorously established to be a quantum spin liquid, and the fractionalization of spins into itinerant Majorana fermions and static $\mathbb{Z}_{2}$ gauge fluxes can be described exactly~\cite{Baskaran-Mandal-2007,Knolle-2014,Knolle-2015,Yoshitake-Nasu-Motome-2016,Yoshitake-Nasu-2017,Nasu-Motome-2021}.
The concepts of spin fractionalization and gauge-theoretical descriptions had been proposed in the context of frustrated antiferromagnetic Heisenberg models even before the Kitaev model~\cite{Arovas-Auerbach-1998,Read-Sachdev-1991,Sachdev-Read-1991,Sachdev-1992,Wang-Vishwanath-2006}.
The Kitaev model, however, provides a concrete platform in which these ideas can be followed without approximation~\cite{Hermanns-Kimchi-2018,Motome-Nasu-2020,Trebst-Hickey-2022}.
This exactly solvable point is also important as a starting point for systematically discussing the effects of perturbations that are unavoidable in real materials, such as magnetic fields and non-Kitaev interactions.
In particular, although a uniform magnetic field breaks time-reversal symmetry and destroys the exact solvability of the model, a low-energy effective theory can still be constructed in the weak-field regime.
For fields applied close to the $[111]$ direction, where the field couples equivalently to the three spin components, an effective spin-chirality interaction is obtained by perturbation theory in the small-field limit.
The Majorana spectrum then becomes gapped, and a topologically nontrivial band structure is realized.
This phase hosts a chiral Majorana edge mode and Ising-type non-Abelian anyonic excitations, and is understood as a chiral spin liquid~\cite{Kitaev-2006}.
The material motivation for the $S=1/2$ problem has also grown rapidly since the Jackeli-Khaliullin mechanism and subsequent studies of honeycomb iridates and $\alpha$-RuCl$_3$~\cite{Jackeli-Khauliullin-2009,Chaloupka-2010,Singh-Gegenwart-2010,Comin-Levy-2012,Sohn-2013,Chaloupka-2013,Foyevtsova-2013,Katukuri-2014,Yamaji-2014,Plumb-2014,Chun-2015,Kubota-2015,Sinn-2016,Winter-2016,Yadav-2016,Rau-Lee-Kee-2016,Winter-2017,Takagi-2019,Haraguchi-2018,Haraguchi-2020,Jang-2021,Takegami-2025}.

While studies of $S=1/2$ Kitaev magnets have progressed both theoretically and experimentally, theoretical understanding of Kitaev magnetism with a general spin quantum number $S$ remains much less developed~\cite{Baskaran-Sen-Shankar-2008,Minakawa-2019,Koga-2020,Hickey-2020,Lee-Kawashima-Kim-2020,Dong-Sheng-2020,Zhu-2020,Khait-Stavropoulos-2021,Chen-2022,Bradley-2022,Fukui-Kato-Nasu-Motome-2022,Georgiou-2024}.
For spin-$1$ and spin-$S$ Kitaev models, numerical studies have clarified ground-state properties, finite-temperature signatures, spin transport, and magnetic-field responses~\cite{Oitmaa-2018,Koga-2018,Minakawa-2019,Koga-2020,Zhu-2020}, and connections between spin-$1/2$ and higher-spin Kitaev physics have been explored through bilayer and multilayer constructions~\cite{Tomishige-Nasu-Koga-2018,Tomishige-Nasu-Koga-2019,Merino-Ralko-2025}.
In this context, an important theoretical advance for higher-$S$ Kitaev models is the clarification of an even-odd effect in the statistics of $\mathbb{Z}_{2}$ gauge charges~\cite{Ma-2023}.
Using a Majorana parton construction for general $S$, it was shown that the local plaquette conserved quantities of the spin-$S$ Kitaev model can be exactly identified as $\mathbb{Z}_{2}$ gauge fluxes, even though the higher-spin model is not exactly solvable.
The corresponding gauge charge~\footnote{In Ref.~\cite{Ma-2023}, the gauge charge defined here is referred to as a giant parton.} is formed as a composite of Majorana fermion components, and its statistics depend on the spin magnitude.
It is fermionic for half-integer spins, for which a deconfined $\mathbb{Z}_{2}$ gauge structure follows, whereas it is bosonic for integer spins and can, in principle, condense.
Accordingly, the case of $S=1$ is not guaranteed by this exact flux structure alone to realize a nontrivial spin-liquid ground state, but it naturally involves bosonic gauge-charge degrees of freedom.
The resulting description in terms of a bosonic parton carrying the $\mathbb{Z}_{2}$ gauge charge is structurally similar to Schwinger-boson descriptions of quantum spin liquids.
This similarity therefore suggests that the Schwinger-boson approach is a natural framework for describing integer-spin Kitaev quantum spin liquids and their instabilities~\cite{Arovas-Auerbach-1998,Read-Sachdev-1991,Sachdev-Read-1991,Sachdev-1992,Flint-Coleman-2009}.
These observations make it particularly intriguing to ask whether a time-reversal-symmetry-breaking spin-liquid regime analogous to that in the $S=1/2$ case can also appear in the $S=1$ Kitaev system.
For $S=1/2$, the zero-field model provides an exactly solvable starting point, and the chiral spin liquid in a weak magnetic field can be understood perturbatively around that limit.
For $S=1$, by contrast, even with the extended Majorana-fermion representation introduced in previous work, it is difficult to identify an unperturbed Hamiltonian that plays a similarly controlled role.
As a result, the existence of a chiral spin-liquid-like state in the $S=1$ Kitaev model cannot be argued for by the same logic as in the $S=1/2$ case.

Against this background, two preceding studies based on Schwinger-boson mean-field theory (SBMFT) are especially relevant. 
One study introduced distinct time-reversal-symmetry-breaking singlet- and triplet-pairing bosonic mean-field Ans\"atze with Wilson phases $\phi_s=\pi/2$ and $\phi_t=\pi/2$, respectively, for the $S=1$ Kitaev model~\cite{Ralko-Merino-2024}.
Among the six bosonic Ans\"atze whose dynamical and static spin structure factors were evaluated by Wick contraction in that study, the triplet $\phi_t=0$ and $\phi_t=\pi/2$ sectors are especially relevant here because they allow a direct comparison between time-reversal-symmetric and time-reversal-breaking flux patterns within the same triplet-pairing channel.
The triplet $\phi_t=\pi/2$ Ansatz exhibited an almost flat low-frequency response, while the singlet $\phi_s=\pi/2$ Ansatz was found to be the most compatible with the exact-diagonalization spectra.
At the pure Kitaev point, the singlet and triplet $\pi/2$-flux solutions were exactly degenerate within the mean-field decoupling in the previous study~\cite{Ralko-Merino-2024}.
Another study focused in particular on the time-reversal-symmetric triplet $\phi_t=0$ ($0$-flux) Ansatz and proposed an evaluation scheme for spin correlations that is inequivalent to the conventional Wick decomposition in SBMFT. In bond-dependent systems such as the Kitaev model, it showed that the two schemes can yield qualitatively different dynamical structure factors~\cite{Sasamoto-Nasu-2025}.
It also evaluated the pure-Kitaev triplet $\pi/2$-flux spectra with both correlation prescriptions.

On the theoretical side, the microscopic mechanism proposed for Ni-based higher-spin Kitaev materials naturally gives an antiferromagnetic sign of the bond-directional exchange, and subsequent spin-$1$ Kitaev studies have emphasized this antiferromagnetic setting~\cite{Stavropoulos-2019,Hickey-2020}.
On the experimental side, neutron-scattering analyses of Na$_2$Ni$_2$TeO$_6$ reported parameter sets on the antiferromagnetic-Kitaev side, while magnetization-plateau measurements on Na$_3$Ni$_2$BiO$_6$ were analyzed in terms of an effective spin-$1$ Hamiltonian that includes antiferromagnetic Kitaev exchange~\cite{Samarakoon-2021,Shangguan-2023}.
Na$_3$Ni$_2$BiO$_6$ has also attracted attention through its honeycomb structure and low-energy NMR response~\cite{Seibel-2013,Shi-2024}, and the spin-wave spectrum of zigzag-ordered KNiAsO$_4$ has been described using an extended Kitaev Hamiltonian~\cite{Taddei-2023}.
At the same time, first-principles studies indicate that the microscopic strength of the Kitaev interaction in Ni-based spin-$1$ honeycomb materials remains unsettled~\cite{Zhao-Li-Hou-2025,Chen-Zhang-Zhu-2026}.
We therefore focus on antiferromagnetic Kitaev exchange while treating the model below as a minimal, rather than material-specific, description.

In this paper, we use the triplet $\pi/2$-flux Ansatz to investigate how weak symmetric off-diagonal exchanges $\Gamma$ and $\Gamma^{\prime}$ affect the spin dynamics of the antiferromagnetic $S=1$ Kitaev model. At the pure Kitaev point, we compare the triplet $\pi/2$-flux Ansatz with its $0$-flux counterpart and use real-space analysis and short-range truncation tests to examine how spectral flatness is related to the range of spin correlations.
We use the triplet $\phi_t=0$ and $\phi_t=\pi/2$ phase patterns considered in Ref.~\cite{Ralko-Merino-2024} together with the bond-operator mean-field decomposition and spin-correlation prescription of Ref.~\cite{Sasamoto-Nasu-2025}.
In the following, ``chiral bosonic Ansatz'' refers specifically to the triplet-pairing $\pi/2$-flux Schwinger-boson mean-field Ansatz, which breaks time-reversal symmetry at the mean-field level.
We restrict the present analysis to the triplet $0$- and $\pi/2$-flux Ans\"atze.
At the pure Kitaev point, this restriction should not be interpreted as an energetic selection of the triplet $\pi/2$-flux Ansatz over its exactly degenerate singlet partner within SBMFT~\cite{Ralko-Merino-2024}.
We then examine weak same-sign perturbations on the line $\Gamma^{\prime}=\Gamma$ within the constrained triplet $\pi/2$-flux manifold and compare the ground-state-energy variation with exact diagonalization (ED) on a 12-site cluster.
We also provide an explicit implementation of the Klein duality within the present bond-operator convention, clarifying how the relevant triplet channels transform under the four-sublattice spin rotation that maps the antiferromagnetic pure Kitaev model onto its ferromagnetic counterpart~\cite{Chaloupka-2010,Kimchi-Vishwanath-2014}.
Our calculations yield two principal findings.
First, at the pure Kitaev point, the triplet $\pi/2$-flux Ansatz yields a flatter dynamical response and more strongly suppresses longer-distance real-space spin correlations than the $0$-flux Ansatz, thereby linking spectral flatness to the short-range correlation structure characteristic of the Kitaev limit.
The dominant low-energy peaks in the $N=24$ ED spectra are also better reproduced by the triplet $\pi/2$-flux Ansatz.
Spectral flatness and short-ranged equal-time spin correlations are, however, not time-reversal-odd observables.
They provide evidence for consistency with Kitaev-like locality but do not by themselves establish microscopic chirality.
Second, weak same-sign exchanges on the line $\Gamma^{\prime}=\Gamma$ admit gapped solutions within the constrained triplet $\pi/2$-flux manifold. The weak ground-state-energy variation near the pure Kitaev point is consistent with the $N=12$ ED results.
Taken together, these results identify the triplet $\pi/2$-flux Ansatz as a physically useful bosonic mean-field description of spin dynamics near the antiferromagnetic $S=1$ Kitaev limit, particularly when weak off-diagonal exchanges are present.

The remainder of this paper is organized as follows.
In Sec.~\ref{sec:Model}, we introduce the $S=1$ Kitaev-$\Gamma$-$\Gamma^{\prime}$ model, specify the parameter line studied, and review the four-sublattice transformation of the pure Kitaev model.
In Sec.~\ref{sec:Method}, we summarize the Schwinger-boson bond-operator formulation, the bosonic Bogoliubov diagonalization, and the calculation of the spin structure factor used throughout this work.
In Sec.~\ref{sec:results}, we present the mean-field Ans\"atze and the resulting dynamical spin structure factors, first for the pure Kitaev model and then for the weak Kitaev-$\Gamma$-$\Gamma^{\prime}$ model.
In Sec.~\ref{sec:Discussion}, we connect real-space correlations to spectral flatness, examine mean-field chirality using three-spin correlations, and compare the ground-state energy and pure-Kitaev spectra with ED.
Section~\ref{sec:Summary} summarizes the conclusions.

\section{Model}
\label{sec:Model}

\begin{figure}[t]
  \centering
      \includegraphics[width=\columnwidth,clip]{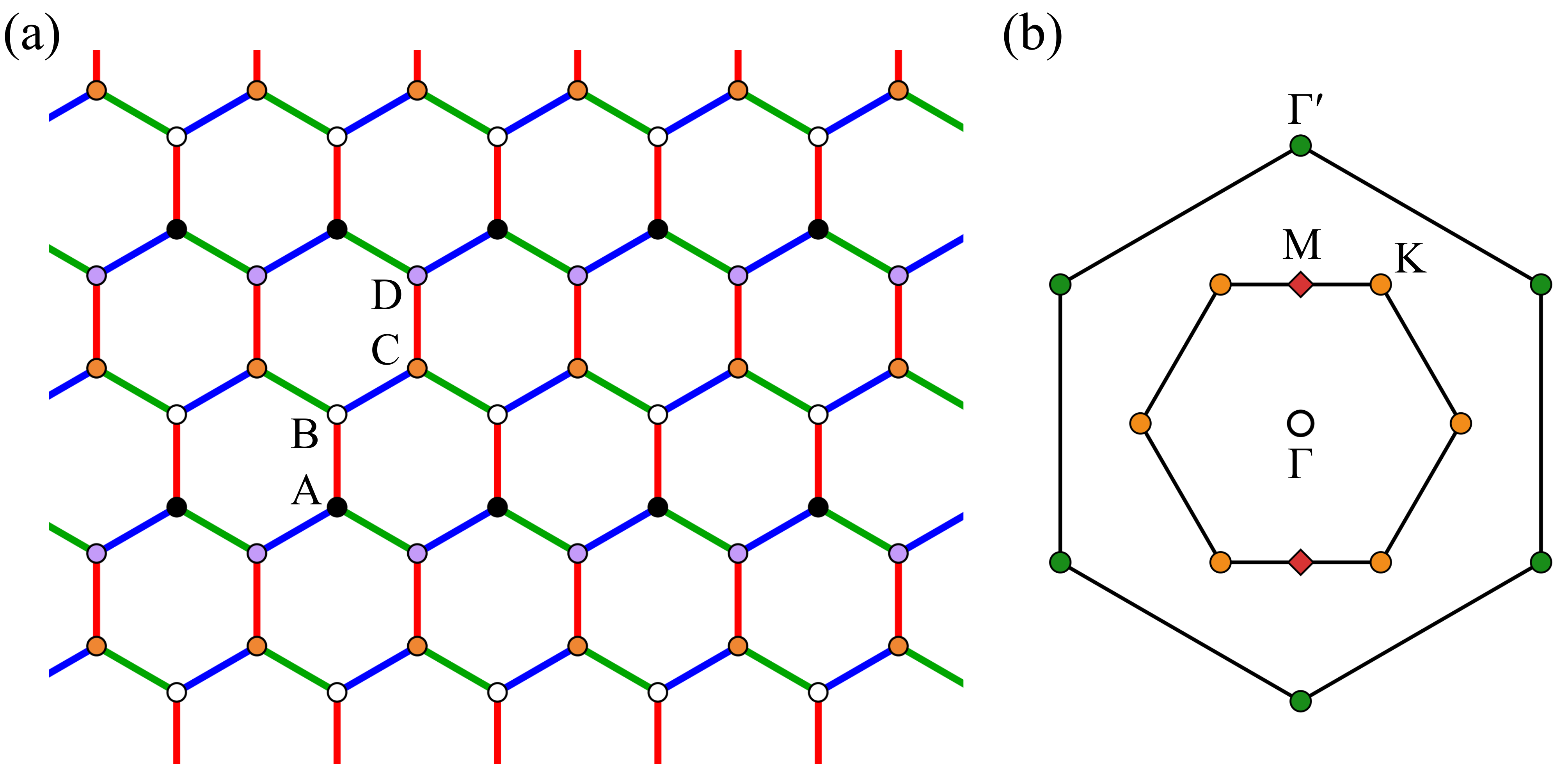}
      \caption{
        (a)
        Schematic picture of the honeycomb lattice on which the $S=1$ Kitaev model is defined.
        The blue, green, and red lines denote the $x$, $y$, and $z$ bonds, respectively.
        The black, white, orange, and purple sites denote the four sublattices A, B, C, and D used in the four-sublattice spin rotation of the pure Kitaev model.
        (b)
        First and extended Brillouin zones of the original honeycomb lattice.
        Filled symbols denote the high-symmetry points.
      }
      \label{fig:honeycomb_lattice}
\end{figure}

We consider the $S=1$ Kitaev-$\Gamma$-$\Gamma^{\prime}$ model on the honeycomb lattice.
The Hamiltonian is given by 
\begin{align}
    \label{eq:Kitaev-Gamma-Gammap-model}
    \mathcal{H}
    &=K\sum_{\langle i,j\rangle_{\gamma}}S_{i}^{\gamma}S_{j}^{\gamma}
    +\Gamma\sum_{\langle i,j\rangle_{\gamma}}\left(S_{i}^{\alpha}S_{j}^{\beta}+S_{i}^{\beta}S_{j}^{\alpha}\right)\notag\\
    &\quad
    +\Gamma^{\prime}\sum_{\langle i,j\rangle_{\gamma}}
    \left(S_{i}^{\gamma}S_{j}^{\alpha}+S_{i}^{\gamma}S_{j}^{\beta}
    +S_{i}^{\alpha}S_{j}^{\gamma}+S_{i}^{\beta}S_{j}^{\gamma}\right),
\end{align}
where $S_{i}^{\gamma}$ $(\gamma=x,y,z)$ denotes the $S=1$ spin operator at site $i$.
The notation $\langle i,j\rangle_{\gamma}$ denotes a nearest-neighbor bond of type $\gamma=x,y,z$ [see Fig.~\ref{fig:honeycomb_lattice}(a)], and $(\alpha,\beta,\gamma)$ is a cyclic permutation of $(x,y,z)$ for each $\gamma$ bond.
The first term is the bond-dependent Kitaev interaction, while the $\Gamma$ and $\Gamma^{\prime}$ terms are the two nearest-neighbor symmetric off-diagonal exchanges commonly considered in extended Kitaev models~\cite{Rau-Lee-Kee-2014,Rousochatzakis-Perkins-Luo-Kee-2024}.
The sign of $K$ determines the magnetic nature of the model: $K>0$ corresponds to the antiferromagnetic (AFM) Kitaev model, whereas $K<0$ corresponds to the ferromagnetic (FM) Kitaev model.
Throughout this paper, we measure energies in units of $\left|K\right|$ and set the length of the primitive translation vectors of the honeycomb lattice to unity.
Accordingly, the calculations with finite off-diagonal exchange are performed for $K>0$.
For these calculations, we restrict ourselves to the same-sign line $\Gamma^{\prime}=\Gamma$ and focus on $\Gamma/|K|=\Gamma^{\prime}/|K|=-0.025$, $0$, and $0.025$.
This choice is not intended as a material-specific parameter set.
Neutron-scattering analyses of Na$_2$Ni$_2$TeO$_6$ motivate only this relative choice.
An optimized solution on the AFM-Kitaev side of an extended Kitaev-Heisenberg model gives $\Gamma=-0.63$ meV and $\Gamma^{\prime}=-0.65$ meV, namely off-diagonal exchanges of comparable magnitude and the same sign~\cite{Samarakoon-2021}.
We do not use this fit to set their magnitudes relative to $K$.
The small magnitudes used here are chosen to keep the analysis close to the pure Kitaev limit and, within the $\pi/2$-flux SBMFT solution studied below, preserve a finite spinon gap.
We therefore regard the calculation as a weak-coupling probe of how same-sign off-diagonal exchanges modify the selected chiral bosonic Ansatz.
The Hamiltonian in Eq.~\eqref{eq:Kitaev-Gamma-Gammap-model} is thus used as a minimal extension of the antiferromagnetic $S=1$ Kitaev model.
Heisenberg exchange and single-ion anisotropy, which may be relevant to specific materials, are not included.

In the pure Kitaev limit, the FM and AFM Kitaev models are connected by a sublattice-dependent rotation of the spin quantization axes, i.e., the Klein duality~\cite{Chaloupka-2010,Kimchi-Vishwanath-2014}.
This is a discrete unitary equivalence of the pure Kitaev Hamiltonian, not a symmetry of the full Kitaev-$\Gamma$-$\Gamma^{\prime}$ model.
To describe it explicitly, we label the four sublattices in Fig.~\ref{fig:honeycomb_lattice}(a) by $\Lambda=\mathrm{A},\mathrm{B},\mathrm{C},\mathrm{D}$ and define
\begin{align}
\label{eq:four-sublattice-unitary}
    U_{\mathrm{A}}=1,\quad
    U_{\mathrm{B}}=e^{-i\pi S^{x}},\quad
    U_{\mathrm{C}}=e^{-i\pi S^{y}},\quad
    U_{\mathrm{D}}=e^{-i\pi S^{z}}.
\end{align}
For a site $i$ belonging to sublattice $\Lambda$, the transformed spin operator is
\begin{align}
\label{eq:spin transformation}
    \widetilde{\bm{S}}_{i}=U_{\Lambda}\bm{S}_{i}U_{\Lambda}^{\dagger}.
\end{align}
Equivalently, $\widetilde{S}_{i}^{\gamma}=\eta_{\Lambda}^{\gamma}S_{i}^{\gamma}$ with
\begin{align}
(\eta_{\mathrm{A}}^{x},\eta_{\mathrm{A}}^{y},\eta_{\mathrm{A}}^{z})&=(+,+,+),&
(\eta_{\mathrm{B}}^{x},\eta_{\mathrm{B}}^{y},\eta_{\mathrm{B}}^{z})&=(+,-,-),\notag\\
(\eta_{\mathrm{C}}^{x},\eta_{\mathrm{C}}^{y},\eta_{\mathrm{C}}^{z})&=(-,+,-),&
(\eta_{\mathrm{D}}^{x},\eta_{\mathrm{D}}^{y},\eta_{\mathrm{D}}^{z})&=(-,-,+).
\end{align}
For the four-sublattice coloring in Fig.~\ref{fig:honeycomb_lattice}(a), every $\gamma$ bond connects two sublattices satisfying $\eta_{\Lambda_i}^{\gamma}\eta_{\Lambda_j}^{\gamma}=-1$.
Thus, for the pure Kitaev Hamiltonian,
\begin{align}
\label{eq:four-sublattice-K-sign}
K\sum_{\langle i,j\rangle_{\gamma}}
\widetilde{S}_{i}^{\gamma}\widetilde{S}_{j}^{\gamma}
=
-K\sum_{\langle i,j\rangle_{\gamma}}
S_{i}^{\gamma}S_{j}^{\gamma}.
\end{align}
This transformation maps the AFM pure Kitaev model with $K>0$ onto the FM one with $K<0$.
The off-diagonal interactions $\Gamma$ and $\Gamma^{\prime}$ do not share this simple sign-changing equivalence in general.
Therefore, the finite-$\Gamma,\Gamma^{\prime}$ calculations below are carried out directly for the AFM case with $K>0$.
The enlarged unit cell used in the SBMFT calculation is chosen to be compatible with this four-sublattice structure and with the flux pattern of the mean-field Ansatz specified later in Sec.~\ref{sec:Mean-field ansatz}.
For theoretical completeness, we also verify how this four-sublattice mapping is represented within SBMFT.
In Appendix~\ref{app:four-sublattice-transformation}, we show how the Schwinger-boson spinors and the triplet bond operators transform in the pure Kitaev limit, and we provide the corresponding FM spectra obtained in this bond-operator convention.

\section{Method}
\label{sec:Method}

\subsection{Schwinger boson theory}
\label{sec:Schwinger boson theory}

Having introduced the microscopic model and parameter regime in Sec.~\ref{sec:Model}, we now formulate the Schwinger boson mean-field approach used to analyze it.
Schwinger boson mean-field theory has been widely used as a bosonic parton framework for addressing frustrated quantum magnets, spin liquids, and their magnetic instabilities~\cite{Arovas-Auerbach-1998,Read-Sachdev-1991,Sachdev-Read-1991,Sachdev-1992,Wang-Vishwanath-2006,Flint-Coleman-2009,Wen-2002}.
It has been applied to a broad range of frustrated Heisenberg and anisotropic spin models on triangular, kagome, square, honeycomb, and related lattices~\cite{Gazza-1993,Manuel-Trumper-1994,Mattsson-Frojdh-1994,Mattsson-1995,Lauchli-2006,Li-Su-Shen-2007,Li-2009,Wang-2010,Mezio-2011,Fak-2012,Merino-Holt-Powell-2014,Halimeh-Punk-2016,Gonzalez-2017,Ghioldi-Gonzalez-2018,Zhang-Ghioldi-2019,Gonzalez-2020,Ghioldi-Zhang-2022,Zhang-2025}.
We first review the Schwinger boson formulation of the SU(2)-symmetric antiferromagnetic Heisenberg model that has been widely used in the literature. We then introduce an extended formulation that can treat anisotropic, SU(2)-breaking interactions, such as Ising-type and Dzyaloshinskii--Moriya interactions.

In the Schwinger boson approach, the spin operator $\bm{S}_{i}$ is represented in terms of two bosonic operators $b_{i\uparrow}$ and $b_{i\downarrow}$ as~\cite{Arovas-Auerbach-1998,Read-Sachdev-1991,Sachdev-Read-1991,Sachdev-1992}
\begin{align}
    \label{eq:Schwinger-boson-representation}
    S_{i}^{\gamma}=\frac{1}{2}\sum_{\mu,\nu}b_{i\mu}^{\dagger}\sigma^{\gamma}_{\mu\nu}b_{i\nu},
\end{align}
where $i$ labels the lattice site on which the spin operator is defined, and $\sigma^{\gamma}$ is the Pauli matrix associated with the spin component $\gamma=x,y,z$.
In general, such a rewriting enlarges the Hilbert space and hence introduces unphysical states.
To faithfully represent the original spin operators, one must impose the local constraint
\begin{align}
    \label{eq:local-constraint}
    n_{i}\equiv \sum_{\mu}b_{i\mu}^{\dagger}b_{i\mu}=2S,
\end{align}
where $S$ denotes the magnitude of the spin.
Physically meaningful values of $S$ are restricted to $S=1/2,1,3/2,\cdots$.
In the Schwinger boson approach, the information on the spin magnitude is encoded in this constraint, which allows a straightforward extension to general spin-$S$.

In what follows, we explain how spin--spin interactions can be expressed within the Schwinger boson formalism.
We begin with the Heisenberg interaction, which has been the standard setting in previous applications of this approach.
Rewriting the Heisenberg interaction using Eq.~\eqref{eq:Schwinger-boson-representation}, one obtains
\begin{align}
    \label{eq:Heisenberg-interaction}
    \bm{S}_{i}\cdot\bm{S}_{j}=\frac{1}{4}\sum_{\gamma=x,y,z}\sum_{\mu,\nu,\rho,\lambda}\sigma^{\gamma}_{\mu\nu}\sigma^{\gamma}_{\rho\lambda}b_{i\mu}^{\dagger}b_{i\nu}b_{j\rho}^{\dagger}b_{j\lambda}.
\end{align}
Using the identity
\begin{align}
    \label{eq:Identity-Heisenberg}
\sum_{\gamma=x,y,z}\sigma^{\gamma}_{\mu\nu}\sigma^{\gamma}_{\rho\lambda}=\sigma^{0}_{\mu\lambda}\sigma_{\nu\rho}^{0}+\sigma_{\mu\rho}^{y}\sigma_{\nu\lambda}^{y},
\end{align}
Eq.~\eqref{eq:Heisenberg-interaction} can be rewritten as
\begin{align}
    \label{eq:Heisenberg-interaction-Schwinger-boson}
    \bm{S}_{i}\cdot\bm{S}_{j}=\normord{\mathcal{B}_{ij}^{\dagger}\mathcal{B}_{ij}}-\mathcal{A}_{ij}^{\dagger}\mathcal{A}_{ij}.
\end{align}
Here we have defined the following SU(2)-invariant bond operators:
\begin{align}
    \mathcal{B}_{ij}
    &=\frac{1}{2}\sum_{\mu,\nu}b_{i\mu}^{\dagger}\sigma_{\mu\nu}^{0}b_{j\nu}=\frac{1}{2}\left(b_{i\uparrow}^{\dagger}b_{j\uparrow}+b_{i\downarrow}^{\dagger}b_{j\downarrow}\right),
    \label{eq:Bond-operator-B}\\
    \mathcal{A}_{ij}
    &=\frac{i}{2}\sum_{\mu,\nu}b_{i\mu}\sigma_{\mu\nu}^{y}b_{j\nu}=\frac{1}{2}\left(b_{i\uparrow}b_{j\downarrow}-b_{i\downarrow}b_{j\uparrow}\right).
    \label{eq:Bond-operator-A}
\end{align}
The operators $\mathcal{B}_{ij}$ and $\mathcal{A}_{ij}$ are referred to as SU(2)-invariant bond operators, since their operator forms remain unchanged under global SU(2) transformations of the Schwinger bosons~\cite{Arovas-Auerbach-1998}.
Their physical meanings correspond to Schwinger boson hopping and spin-singlet resonance, respectively.
Because the Heisenberg interaction is SU(2) symmetric, it is natural that it can be expressed in terms of SU(2)-invariant bond operators as in Eq.~\eqref{eq:Heisenberg-interaction-Schwinger-boson}.
Here, $\normord{\mathcal{O}_{1}\mathcal{O}_{2}}$ denotes normal ordering, in which all creation operators are arranged to the left of annihilation operators.

On the other hand, a generic spin--spin interaction is not necessarily SU($2$) symmetric, and thus the SU($2$)-invariant bond operators need to be extended in order to represent such interactions.
Such extensions are natural in the present context because Kitaev materials and related spin-orbit-coupled magnets are commonly described by exchange Hamiltonians containing bond-dependent Ising terms, off-diagonal symmetric anisotropies, and other SU(2)-breaking interactions~\cite{Rau-Lee-Kee-2014,Rau-Lee-Kee-2016,Winter-2016,Winter-2017,Rousochatzakis-Perkins-Luo-Kee-2024}.
To this end, we introduce the following six SU($2$)-breaking bond operators:
\begin{align}
    \mathcal{C}_{ij}^{\gamma}
    &=\frac{1}{2}\sum_{\mu,\nu}b_{i\mu}^{\dagger}\sigma_{\mu\nu}^{\gamma}b_{j\nu},
    \label{eq:Bond-operator-C}\\
    \mathcal{D}_{ij}^{\gamma}
    &=\frac{i}{2}\sum_{\mu,\nu}b_{i\mu}\left(\sigma^{y}\sigma^{\gamma}\right)_{\mu\nu}b_{j\nu}.
    \label{eq:Bond-operator-D}
\end{align}
Using these operators, we show how SU(2)-nonsymmetric spin interactions can be described, focusing on the Ising-type interaction and the off-diagonal $\Gamma$-type interaction employed in this work.
Substituting Eq.~\eqref{eq:Schwinger-boson-representation} into the Ising-type interaction $S_{i}^{\gamma}S_{j}^{\gamma}$, we obtain
\begin{align}
    \label{eq:Ising-interaction}
    S_{i}^{\gamma}S_{j}^{\gamma}=\frac{1}{4}\sum_{\mu,\nu,\rho,\lambda}\sigma^{\gamma}_{\mu\nu}\sigma^{\gamma}_{\rho\lambda}b_{i\mu}^{\dagger}b_{i\nu}b_{j\rho}^{\dagger}b_{j\lambda}.
\end{align}
Using the identities
\begin{align}
    \label{eq:identity-Ising}
    \sigma_{\mu\nu}^{x}\sigma_{\rho\lambda}^{x}
    &=\frac{1}{2}\left(\sigma_{\mu\lambda}^{0}\sigma_{\nu\rho}^{0}-\sigma_{\mu\rho}^{z}\sigma_{\nu\lambda}^{z}+\sigma_{\mu\lambda}^{x}\sigma_{\nu\rho}^{x}+\sigma_{\mu\rho}^{y}\sigma_{\nu\lambda}^{y}\right),\\
    \sigma_{\mu\nu}^{y}\sigma_{\rho\lambda}^{y}
    &=\frac{1}{2}\left(\sigma_{\mu\lambda}^{0}\sigma_{\nu\rho}^{0}-\sigma_{\mu\rho}^{0}\sigma_{\nu\lambda}^{0}-\sigma_{\mu\lambda}^{y}\sigma_{\nu\rho}^{y}+\sigma_{\mu\rho}^{y}\sigma_{\nu\lambda}^{y}\right),\\
    \sigma_{\mu\nu}^{z}\sigma_{\rho\lambda}^{z}
    &=\frac{1}{2}\left(\sigma_{\mu\lambda}^{0}\sigma_{\nu\rho}^{0}-\sigma_{\mu\rho}^{x}\sigma_{\nu\lambda}^{x}+\sigma_{\mu\lambda}^{z}\sigma_{\nu\rho}^{z}+\sigma_{\mu\rho}^{y}\sigma_{\nu\lambda}^{y}\right),
\end{align}
Eq.~\eqref{eq:Ising-interaction} can be rewritten as
\begin{align}
    \label{eq:Ising-interaction-Schwinger-boson}
    S_{i}^{\gamma}S_{j}^{\gamma}=\frac{1}{2}\left(\normord{\mathcal{B}_{ij}^{\dagger}\mathcal{B}_{ij}}-\mathcal{D}_{ij}^{\gamma\dagger}\mathcal{D}_{ij}^{\gamma}+\normord{\mathcal{C}_{ij}^{\gamma\dagger}\mathcal{C}_{ij}^{\gamma}}-\mathcal{A}_{ij}^{\dagger}\mathcal{A}_{ij}\right).
\end{align}
Here, we notice that there exist operator identities that hold identically among $\mathcal{A}_{ij}$, $\mathcal{B}_{ij}$, $\mathcal{C}_{ij}^{\gamma}$, and $\mathcal{D}_{ij}^{\gamma}$.

\begin{align}
    \label{eq:relationship-Ising-Heisenberg}
    &\bm{S}_{i}\cdot\bm{S}_{j}=\sum_{\gamma=x,y,z}S_{i}^{\gamma}S_{j}^{\gamma}\notag\\
    &=\frac{3}{2}\left(\normord{\mathcal{B}_{ij}^{\dagger}\mathcal{B}_{ij}}-\mathcal{A}_{ij}^{\dagger}\mathcal{A}_{ij}\right)+\frac{1}{2}\sum_{\gamma=x,y,z}\left(\normord{\mathcal{C}_{ij}^{\gamma\dagger}\mathcal{C}_{ij}^{\gamma}}-\mathcal{D}_{ij}^{\gamma\dagger}\mathcal{D}_{ij}^{\gamma}\right).
\end{align}
By applying Eq.~\eqref{eq:Heisenberg-interaction-Schwinger-boson} to Eq.~\eqref{eq:relationship-Ising-Heisenberg}, we find that the following operator identity holds:
\begin{align}
    \label{eq:constraint-Bond-operator}
    \normord{\mathcal{B}_{ij}^{\dagger}\mathcal{B}_{ij}}-\mathcal{A}_{ij}^{\dagger}\mathcal{A}_{ij}=\sum_{\gamma=x,y,z}\left(\mathcal{D}_{ij}^{\gamma\dagger}\mathcal{D}_{ij}^{\gamma}-\normord{\mathcal{C}_{ij}^{\gamma\dagger}\mathcal{C}_{ij}^{\gamma}}\right).
\end{align}
Using this constraint, we see that $S_{i}^{\gamma}S_{j}^{\gamma}$ admits the following two equivalent representations:
\begin{subequations}
\label{eq:Heisenberg-interaction-Schwinger-boson-ver2}
\begin{align}
    S_{i}^{\gamma}S_{j}^{\gamma}
    &=\dfrac{1}{2}\left(\normord{\mathcal{B}_{ij}^{\dagger}\mathcal{B}_{ij}}-\mathcal{D}_{ij}^{\gamma\dagger}\mathcal{D}_{ij}^{\gamma}+\normord{\mathcal{C}_{ij}^{\gamma\dagger}\mathcal{C}_{ij}^{\gamma}}-\mathcal{A}_{ij}^{\dagger}\mathcal{A}_{ij}\right),
    \label{eq:Ising-representation-upper}\\
    S_{i}^{\gamma}S_{j}^{\gamma}
    &=\frac{1}{2}\sum_{\mu\neq\gamma}\left(\mathcal{D}_{ij}^{\mu\dagger}\mathcal{D}_{ij}^{\mu}-\normord{\mathcal{C}_{ij}^{\mu\dagger}\mathcal{C}_{ij}^{\mu}}\right).
    \label{eq:Ising-representation-lower}
\end{align}
\end{subequations}
The choice between these two equivalent representations should be made on the basis of physical considerations.
In the present work, for the antiferromagnetic Kitaev model of interest, we carry out our analysis using Eq.~\eqref{eq:Ising-representation-upper} under the physical considerations discussed in Sec.~\ref{sec:Mean-field ansatz}.

Furthermore, for off-diagonal spin interactions of the form $S_{i}^{\alpha}S_{j}^{\beta}\ (\alpha\neq\beta)$, such as the $\Gamma$ and $\Gamma^{\prime}$ interactions studied in this work, substituting Eq.~\eqref{eq:Schwinger-boson-representation} yields
\begin{align}
    \label{eq:off-diagonal-interaction}
    S_{i}^{\alpha}S_{j}^{\beta}=\frac{1}{4}\sum_{\mu,\nu,\rho,\lambda}\sigma^{\alpha}_{\mu\nu}\sigma^{\beta}_{\rho\lambda}b_{i\mu}^{\dagger}b_{i\nu}b_{j\rho}^{\dagger}b_{j\lambda}.
\end{align}
Using the following identities,
\begin{align}  
    \label{eq:off-diagonal-identity}  
    \sigma_{\mu\nu}^{x}\sigma_{\rho\lambda}^{y}  
    &=-\sigma_{\mu\lambda}^{x}\sigma_{\nu\rho}^{y}+\sigma_{\mu\rho}^{x}\sigma_{\nu\lambda}^{y}  
    =\sigma_{\mu\lambda}^{y}\sigma_{\nu\rho}^{x}-\sigma_{\mu\rho}^{y}\sigma_{\nu\lambda}^{x}\nonumber\\  
    &=-i\sigma_{\mu\rho}^{z}\sigma_{\nu\lambda}^{0}+i\sigma_{\mu\lambda}^{z}\sigma_{\nu\rho}^{0}=i\sigma_{\mu\rho}^{0}\sigma_{\nu\lambda}^{z}-i\sigma_{\mu\lambda}^{0}\sigma_{\nu\rho}^{z},\\  
    \sigma_{\mu\nu}^{y}\sigma_{\rho\lambda}^{z}  
    &=\sigma_{\mu\lambda}^{y}\sigma_{\nu\rho}^{z}-\sigma_{\mu\rho}^{z}\sigma_{\nu\lambda}^{y}  
    =-\sigma_{\mu\lambda}^{z}\sigma_{\nu\rho}^{y}+\sigma_{\mu\rho}^{y}\sigma_{\nu\lambda}^{z}\nonumber\\  
    &=-i\sigma_{\mu\rho}^{0}\sigma_{\nu\lambda}^{x}+i\sigma_{\mu\lambda}^{x}\sigma_{\nu\rho}^{0}=i\sigma_{\mu\rho}^{x}\sigma_{\nu\lambda}^{0}-i\sigma_{\mu\lambda}^{0}\sigma_{\nu\rho}^{x},\\  
    \sigma_{\mu\nu}^{z}\sigma_{\rho\lambda}^{x}  
    &=\sigma_{\mu\lambda}^{z}\sigma_{\nu\rho}^{x}+i\sigma_{\mu\rho}^{0}\sigma_{\nu\lambda}^{y}  
    =\sigma_{\mu\lambda}^{x}\sigma_{\nu\rho}^{z}+i\sigma_{\mu\rho}^{y}\sigma_{\nu\lambda}^{0}\nonumber\\  
    &=\sigma_{\mu\rho}^{x}\sigma_{\nu\lambda}^{z}+i\sigma_{\mu\lambda}^{y}\sigma_{\nu\rho}^{0}=\sigma_{\mu\rho}^{z}\sigma_{\nu\lambda}^{x}+i\sigma_{\mu\lambda}^{0}\sigma_{\nu\rho}^{y},  
\end{align}  
we obtain
\begin{align}  
    \label{eq:off-diagonal-interaction-Schwinger-boson-representation}  
    &S_{i}^{\alpha}S_{j}^{\beta}\nonumber\\  
    &=\frac{1}{2}\left[\normord{\mathcal{C}_{ij}^{\alpha\dagger}\mathcal{C}_{ij}^{\beta}}+\normord{\mathcal{C}_{ij}^{\beta\dagger}\mathcal{C}_{ij}^{\alpha}}+i\sum_\gamma \epsilon_{\alpha\beta\gamma}\left(\mathcal{D}_{ij}^{\gamma\dagger}\mathcal{A}_{ij}-\mathcal{A}_{ij}^{\dagger}\mathcal{D}_{ij}^{\gamma}\right)\right]\\  
    &=-\frac{1}{2}\left[\mathcal{D}_{ij}^{\alpha\dagger}\mathcal{D}_{ij}^{\beta}+\mathcal{D}_{ij}^{\beta\dagger}\mathcal{D}_{ij}^{\alpha}+i\sum_\gamma \epsilon_{\alpha\beta\gamma}\left(\normord{\mathcal{C}_{ij}^{\gamma\dagger}\mathcal{B}_{ij}}-\normord{\mathcal{B}_{ij}^{\dagger}\mathcal{C}_{ij}^{\gamma}}\right)\right],  
\end{align}  
where we have used the Hermiticity of the interaction, $\left(S_{i}^{\alpha}S_{j}^{\beta}\right)^{\dagger}=S_{i}^{\alpha}S_{j}^{\beta}$, and the repeated index $\gamma$ in the Levi-Civita terms is summed over $x,y,z$.
As is evident, two alternative representations are also obtained for off-diagonal interactions.
In what follows, we use the symmetrized form of these expressions and describe the $\Gamma$ interaction as a quadratic form of bond operators, as shown below.

\begin{align}
    \label{eq:off-diagonal-interaction-Schwinger-boson}
    S_{i}^{\alpha}S_{j}^{\beta}+S_{i}^{\beta}S_{j}^{\alpha}=\frac{1}{2}\left(\normord{\mathcal{C}_{ij}^{\alpha\dagger}\mathcal{C}_{ij}^{\beta}}+\normord{\mathcal{C}_{ij}^{\beta\dagger}\mathcal{C}_{ij}^{\alpha}}-\mathcal{D}_{ij}^{\alpha\dagger}\mathcal{D}_{ij}^{\beta}-\mathcal{D}_{ij}^{\beta\dagger}\mathcal{D}_{ij}^{\alpha}\right).
\end{align}

As is clear from the discussion above, within the Schwinger boson formalism one can use the bond-operator vector
$\bm{\mathcal{Q}}_{ij}=\left(\mathcal{A}_{ij},\mathcal{B}_{ij},\mathcal{C}_{ij}^{x},\mathcal{C}_{ij}^{y},\mathcal{C}_{ij}^{z},\mathcal{D}_{ij}^{x},\mathcal{D}_{ij}^{y},\mathcal{D}_{ij}^{z}\right)^{T}$
to express a generic spin--spin interaction in a quadratic form of bond operators as
\begin{align}
    S_{i}^{\alpha}S_{j}^{\beta}=\sum_{p,q}A_{pq}^{\alpha\beta}\normord{\mathcal{Q}_{ij}^{p\dagger}\mathcal{Q}_{ij}^{q}},
\end{align}
where $A_{pq}^{\alpha\beta}$ are numerical coefficients.
For example, for the representation of $S_{i}^{z}S_{j}^{z}$, one has $A_{11}^{zz}=-1/2$, $A_{22}^{zz}=+1/2$, $A_{55}^{zz}=+1/2$, and $A_{88}^{zz}=-1/2$, while all other $A_{pq}^{zz}$ vanish.
Since these coefficients depend on the spin components, we explicitly indicate the $(\alpha,\beta)$ dependence as in $A_{pq}^{\alpha\beta}$.

\subsection{Mean-field theory}
\label{sec:Mean-field theory}
In this section, we explain the Schwinger boson mean-field theory (SBMFT), in which the Schwinger boson formalism introduced in Sec.~\ref{sec:Schwinger boson theory} is applied to quantum magnets and analyzed within the mean-field framework.
The mean-field formulation also provides a convenient language for classifying symmetric and symmetry-breaking spin-liquid Ans\"atze through projective symmetry considerations~\cite{Wen-1989,Wen-2002,Wang-Vishwanath-2006,Messio-Lhuillier-2013,Messio-Bieri-2017,Schneider-2022}.
First, we apply the Schwinger boson formalism to the general Hamiltonian for a quantum magnet,
\begin{align}
    \label{eq:magnetic-Hamiltonian}
    \mathcal{H}=\frac{1}{2}\sum_{i,j}\sum_{\alpha,\beta}J_{ij}^{\alpha\beta}S_{i}^{\alpha}S_{j}^{\beta}.
\end{align}
Here, $J_{ij}^{\alpha\beta}$ denotes the exchange coupling acting between the spin operators $S_{i}^{\alpha}$ and $S_{j}^{\beta}$ defined on a lattice with $N$ sites, and it satisfies $J_{ij}^{\alpha\beta}=J_{ji}^{\beta\alpha}$.
As described in Sec.~\ref{sec:Schwinger boson theory}, such a Hamiltonian can be written in the following bilinear form in terms of bond operators:
\begin{align}
    \label{eq:bilinear-form-bond-operator}
    \mathcal{H}=\frac{1}{2}\sum_{i,j}\sum_{\alpha,\beta}J_{ij}^{\alpha\beta}\sum_{p,q}A_{pq}^{\alpha\beta}\normord{\mathcal{Q}_{ij}^{p\dagger}\mathcal{Q}_{ij}^{q}}.
\end{align}
Here, $\mathcal{Q}_{ij}^{p}$ denotes the $p$th component of the bond-operator vector $\bm{\mathcal{Q}}_{ij}$.
We then apply the mean-field approximation to this bilinear representation of the bond operators as follows:
\begin{align}
    \label{eq:bilinear-form-bond-operator-mean-field}
    \mathcal{H}
    &\approx\frac{1}{2}\sum_{i,j}\sum_{\alpha,\beta}J_{ij}^{\alpha\beta}\sum_{p,q}A_{pq}^{\alpha\beta}
    \Bigl(\langle\mathcal{Q}_{ij}^{p\dagger}\rangle\mathcal{Q}_{ij}^{q}
    +\langle\mathcal{Q}_{ij}^{q}\rangle\mathcal{Q}_{ij}^{p\dagger}\notag\\
    &\hspace{41mm}
    -\langle\mathcal{Q}_{ij}^{p\dagger}\rangle\langle\mathcal{Q}_{ij}^{q}\rangle\Bigr).
\end{align}
Furthermore, by introducing the term
\begin{align}
    \label{eq:constraint-mean-field}
    \mathcal{H}_{\text{c}}=\lambda\sum_{i}\left(n_{i}-2S\right)
\end{align}
with the Lagrange multiplier $\lambda$, so as to impose a global relaxation of the local constraint in Eq.~\eqref{eq:local-constraint}, we obtain the mean-field Hamiltonian
\begin{align}
    \label{eq:mean-field-Hamiltonian}
    \mathcal{H}^{\text{MF}}=\sum_{i,j}\mathcal{H}_{ij}+\mathcal{H}_{\text{c}}.
\end{align}
Here, $\mathcal{H}_{ij}$, which depends on the bond $i,j$, is defined as
\begin{align}
    \label{eq:definition-Hij}
    \mathcal{H}_{ij}=\frac{1}{2}\sum_{\alpha,\beta}\sum_{p,q}J_{ij}^{\alpha\beta}A_{pq}^{\alpha\beta}\left(\langle\mathcal{Q}_{ij}^{p\dagger}\rangle\mathcal{Q}_{ij}^{q}+\langle\mathcal{Q}_{ij}^{q}\rangle\mathcal{Q}_{ij}^{p\dagger}-\langle\mathcal{Q}_{ij}^{p\dagger}\rangle\langle\mathcal{Q}_{ij}^{q}\rangle\right).
\end{align}
Although the physical value considered in this work is $S=1$, the spin length $S$ can also be varied continuously within SBMFT as a parameter controlling the strength of quantum fluctuations: a smaller $S$ corresponds to stronger quantum fluctuations.
This provides a standard way to describe how quantum fluctuations melt magnetic order.

Since each component of the bond-operator vector $\bm{\mathcal{Q}}_{ij}$ is a bilinear constructed from Schwinger-boson creation or annihilation operators at sites $i$ and $j$, we introduce the site-local Nambu vector
\begin{align}
    \label{eq:site-Nambu-vector}
    B_i^{\dagger}
    =
    \left(
    b_{i\uparrow}^{\dagger},
    b_{i\downarrow}^{\dagger},
    b_{i\uparrow},
    b_{i\downarrow}
    \right).
\end{align}
The mean-field Hamiltonian can then be written in the bilinear form
\begin{align}
    \label{eq:BdG-real-space}
    \mathcal{H}^{\text{MF}}
    =
    \frac{1}{2}\sum_{i,j}B_i^{\dagger}\mathcal{M}_{ij}B_j
    +\text{const},
\end{align}
where $\mathcal{M}_{ij}$ is a $4\times4$ matrix in the local Nambu and spin space and satisfies $\mathcal{M}_{ij}^{\dagger}=\mathcal{M}_{ji}$.
We label a site as $i=(l,m)$, where $l=1,2,\ldots,N/M$ is the unit-cell index and $m=1,2,\ldots,M$ is the sublattice index, so that $\mathcal{M}_{ij}=\mathcal{M}_{(l,m)(l^{\prime},m^{\prime})}$.
From here on, we omit the additive constant in Eq.~\eqref{eq:BdG-real-space}, which contributes only to the energy shift.
For each sublattice, we define the Fourier-transformed Nambu vector by
\begin{align}
    \label{eq:Nambu-vector-momentum-space}
    B_{\bm{k},m}^{\dagger}
    &=
    \sqrt{\frac{M}{N}}
    \sum_l B_{(l,m)}^{\dagger}e^{i\bm{k}\cdot\bm R_l}
    \notag\\
    &=
    \left(
    b_{\bm{k},m,\uparrow}^{\dagger},
    b_{\bm{k},m,\downarrow}^{\dagger},
    b_{-\bm{k},m,\uparrow},
    b_{-\bm{k},m,\downarrow}
    \right).
\end{align}
The momentum-space Hamiltonian is therefore
\begin{align}
    \label{eq:BdG-momentum-space}
    \mathcal{H}^{\text{MF}}
    =
    \frac{1}{2}
    \sum_{\bm{k}}^{\text{B.Z.}}
    \sum_{m,m^{\prime}}
    B_{\bm{k},m}^{\dagger}
    (\mathcal{M}_{\bm{k}})_{mm^{\prime}}
    B_{\bm{k},m^{\prime}},
\end{align}
where the sum over $\bm{k}$ is taken over the first Brillouin zone of the superlattice and $\bm R_l$ is the representative position of unit cell $l$.
The $4\times4$ block connecting sublattices $m$ and $m^{\prime}$ is
\begin{align}
    \label{eq:definition-Mk}
    (\mathcal{M}_{\bm{k}})_{mm^{\prime}}
    =
    \sum_{l^{\prime}}
    \mathcal{M}_{(l,m)(l^{\prime},m^{\prime})}
    e^{-i\bm{k}\cdot(\bm R_l-\bm R_{l^{\prime}})}.
\end{align}
Thus, $\mathcal M_{\bm k}$ is an $M\times M$ block matrix, with $4\times4$ blocks, and has total dimension $4M\times4M$.
Because the real-space matrix depends only on the relative coordinate $\bm R_l-\bm R_{l^{\prime}}$, $\mathcal M_{\bm k}$ is independent of $l$.

We diagonalize the bosonic BdG Hamiltonian by a Bogoliubov transformation~\cite{Arovas-Auerbach-1998,Sasamoto-Nasu-2025,Sasamoto-Nasu-2026}.
Collecting the particle components of $B_{\bm k,m}$ before the hole components, the corresponding $4M\times4M$ metric is
\begin{align}
    \label{eq:bosonic-metric}
    \Sigma_{3}=
    \begin{pmatrix}
        \bm{1}_{2M\times 2M} & 0\\
        0 & -\bm{1}_{2M\times 2M}
    \end{pmatrix},
\end{align}
which encodes the canonical commutation relations in this ordering.
The Bogoliubov transformation is
\begin{align}
    \label{eq:Bogoliubov-transformation}
    \begin{aligned}
    \Bigl(&
    b_{\bm k,1,\uparrow},\ldots,b_{\bm k,M,\uparrow},
    b_{\bm k,1,\downarrow},\ldots,b_{\bm k,M,\downarrow},\\
    &b_{-\bm k,1,\uparrow}^{\dagger},\ldots,b_{-\bm k,M,\uparrow}^{\dagger},
    b_{-\bm k,1,\downarrow}^{\dagger},\ldots,b_{-\bm k,M,\downarrow}^{\dagger}
    \Bigr)^{T}
    =\mathcal T_{\bm k}\Gamma_{\bm k},
    \end{aligned}
\end{align}
where
\begin{align}
    \label{eq:Gamma-vector}
    \Gamma_{\bm{k}}^{\dagger}
    =
    \Bigl(
    \gamma_{\bm{k},1}^{\dagger},\cdots,\gamma_{\bm{k},2M}^{\dagger},
    \gamma_{-\bm{k},1},\cdots,\gamma_{-\bm{k},2M}
    \Bigr).
\end{align}
Here, $\gamma_{\bm{k},\eta}$ and $\gamma_{\bm{k},\eta}^{\dagger}$ are bosonic quasiparticle annihilation and creation operators.
In order for the transformation in Eq.~\eqref{eq:Bogoliubov-transformation} to preserve the bosonic commutation relations, the matrix $\mathcal{T}_{\bm{k}}$ must be paraunitary:
\begin{align}
    \label{eq:paraunitary-condition}
    \mathcal{T}_{\bm{k}}^{\dagger}\Sigma_{3}\mathcal{T}_{\bm{k}}
    =
    \mathcal{T}_{\bm{k}}\Sigma_{3}\mathcal{T}_{\bm{k}}^{\dagger}
    =
    \Sigma_{3}.
\end{align}
Equivalently, $\mathcal{T}_{\bm{k}}^{-1}=\Sigma_{3}\mathcal{T}_{\bm{k}}^{\dagger}\Sigma_{3}$.
The physical quasiparticle energies are obtained from the generalized eigenvalue problem
\begin{align}
    \label{eq:bosonic-BdG-eigenproblem}
    \Sigma_{3}\mathcal{M}_{\bm{k}}\mathcal{T}_{\bm{k}}
    =
    \mathcal{T}_{\bm{k}}\Sigma_{3}\mathcal{E}_{\bm{k}},
\end{align}
with the normalization in Eq.~\eqref{eq:paraunitary-condition}.
Equivalently, the same transformation diagonalizes the BdG matrix as
\begin{align}
    \label{eq:bosonic-BdG-diagonalization}
    \mathcal{T}_{\bm{k}}^{\dagger}
    \mathcal{M}_{\bm{k}}
    \mathcal{T}_{\bm{k}}
    =
    \mathcal{E}_{\bm{k}}.
\end{align}
For a stable spin-liquid mean-field solution, all positive-norm eigenmodes have positive energies.
We arrange the diagonal matrix as
\begin{align}
    \label{eq:BdG-energy-matrix}
    \mathcal{E}_{\bm{k}}
    =
    \operatorname{diag}
    \left(
    \varepsilon_{\bm{k},1},\cdots,\varepsilon_{\bm{k},2M},
    \varepsilon_{-\bm{k},1},\cdots,\varepsilon_{-\bm{k},2M}
    \right),
\end{align}
where $\varepsilon_{\bm{k},\eta}>0$ denotes the energy of the $\eta$th bosonic spinon band.
With these definitions, the diagonalized Hamiltonian is
\begin{align}
    \label{eq:diagonalized-mean-field-Hamiltonian}
    \mathcal{H}^{\text{MF}}
    =
    \frac{1}{2}
    \sum_{\bm{k}}^{\text{B.Z.}}
    \Gamma_{\bm{k}}^{\dagger}
    \mathcal{E}_{\bm{k}}
    \Gamma_{\bm{k}},
\end{align}
up to the additive constant omitted above.
The thermal occupation number is given by the Bose distribution
\begin{align}
    \label{eq:Bose-distribution}
    n_{\text{B}}(\varepsilon)
    =
    \frac{1}{e^{\beta\varepsilon}-1},
\end{align}
where $\beta=1/T$ is the inverse temperature.
The expectation values of the bond operators $\langle\mathcal{Q}_{ij}^{p}\rangle$ and the average constraint $\sum_{i}\langle n_{i}\rangle=2SN$ are evaluated using Eq.~\eqref{eq:Bogoliubov-transformation} and Eq.~\eqref{eq:Bose-distribution}.
The mean-field parameters and the Lagrange multiplier $\lambda$ are then determined self-consistently.
A vanishing minimum positive eigenvalue signals the loss of the gapped saddle and a possible spinon-condensation instability.

\subsection{Calculation of the spin structure factor}
\label{sec:Calculation of the spin structure factor}

In this section, we explain how the dynamical spin structure factor and the static spin structure factor are calculated from the diagonalized mean-field Hamiltonian.
Spin correlations have been evaluated within SBMFT in many previous studies of frustrated magnets, both for static structure factors and for dynamical responses~\cite{Arovas-Auerbach-1998,Mezio-2011,Mezio-2013,Ghioldi-Gonzalez-2018,Zhang-Ghioldi-2019,Ghioldi-Zhang-2022}.
In the present work, we evaluate spin correlations in the bond-operator basis following Ref.~\cite{Sasamoto-Nasu-2025}.
The dynamical spin structure factor is defined as the Fourier transform of the spin correlation function in time and space:
\begin{align}
    \label{eq:definition-dynamical-spin-structure-factor}
    S^{\alpha\alpha^{\prime}}(\bm{q},\omega)=\int_{-\infty}^{\infty}dt e^{i\omega t}\frac{1}{N}\sum_{i,j}\langle S_{i}^{\alpha}(t)S_{j}^{\alpha^{\prime}}\rangle e^{-i\bm{q}\cdot\left(\bm{r}_{i}-\bm{r}_{j}\right)}.
\end{align}
Here and below in this subsection, $\alpha,\alpha^{\prime}=x,y,z$ denote spin components.
For the actual calculation, we formulate the same response in imaginary time.
We introduce
\begin{align}
    \label{eq:imaginary-time-spin-susceptibility-real-space}
    \chi_{ij}^{\alpha\alpha^{\prime}}(\tau)
    =
    \langle T_{\tau} S_i^\alpha(\tau)S_j^{\alpha^{\prime}}(0)\rangle,
    \qquad
    0\leq \tau < \beta,
\end{align}
where $T_{\tau}$ is the imaginary-time-ordering operator.
The imaginary-time spin operator is written as
\begin{align}
    \label{eq:imaginary-time-spin-operator}
    S_{i}^{\alpha}(\tau)=\frac{1}{2}\sum_{\mu,\nu}\bar{b}_{i\mu}(\tau)\sigma^{\alpha}_{\mu\nu}b_{i\nu}(\tau),
\end{align}
with
\begin{align}
    \label{eq:imaginary-time-Schwinger-boson-operator}
    b_{i\mu}(\tau)
    &=
    e^{\tau\mathcal{H}_{\text{MF}}} b_{i\mu} e^{-\tau\mathcal{H}_{\text{MF}}},\\
    \bar{b}_{i\mu}(\tau)
    &=
    e^{\tau\mathcal{H}_{\text{MF}}} b_{i\mu}^{\dagger} e^{-\tau\mathcal{H}_{\text{MF}}}.
\end{align}
Here, the bar denotes the imaginary-time-evolved creation operator.
Correspondingly, the Nambu operator and its conjugate in imaginary time are
$B_i(\tau)\equiv
e^{\tau\mathcal{H}_{\text{MF}}}B_i e^{-\tau\mathcal{H}_{\text{MF}}}$ and
$\overline{B}_{i}(\tau)\equiv
e^{\tau\mathcal{H}_{\text{MF}}}B_i^{\dagger}e^{-\tau\mathcal{H}_{\text{MF}}}$,
respectively.
The Matsubara susceptibility is
\begin{align}
    \label{eq:Matsubara-spin-susceptibility}
    \chi^{\alpha\alpha^{\prime}}(\bm{q},i\omega_n)
    =
    \frac{1}{N}
    \sum_{i,j}
    e^{-i\bm{q}\cdot(\bm{r}_i-\bm{r}_j)}
    \int_0^{\beta}d\tau\,
    e^{i\omega_n\tau}
    \chi_{ij}^{\alpha\alpha^{\prime}}(\tau),
\end{align}
where $\omega_n=2\pi n/\beta$ is a bosonic Matsubara frequency.
After the analytic continuation $i\omega_n\rightarrow \omega+i\delta$, the dynamical spin structure factor is obtained as
\begin{align}
    \label{eq:DSSF-from-Matsubara-susceptibility}
    S^{\alpha\alpha^{\prime}}(\bm{q},\omega)
    =
    \frac{2}{1-e^{-\beta\omega}}
    \operatorname{Im}
    \chi^{\alpha\alpha^{\prime}}(\bm{q},\omega+i\delta).
\end{align}
The positive infinitesimal $\delta$ is replaced by a finite broadening in numerical calculations.
The equal-time structure factor follows from the frequency integral of the dynamical response,
\begin{align}
    \label{eq:definition-static-spin-structure-factor}
    S^{\alpha\alpha^{\prime}}(\bm{q})
    &=
    \int_{-\infty}^{\infty}\frac{d\omega}{2\pi}
    S^{\alpha\alpha^{\prime}}(\bm{q},\omega)
    \notag\\
    &=
    \frac{1}{N}
    \sum_{i,j}
    \langle S_{i}^{\alpha}S_{j}^{\alpha^{\prime}}\rangle
    e^{-i\bm{q}\cdot\left(\bm{r}_{i}-\bm{r}_{j}\right)}.
\end{align}
Thus, the central task is to evaluate $\chi_{ij}^{\alpha\alpha^{\prime}}(\tau)$ within the approximate mean-field theory.

We follow the prescription of Ref.~\cite{Sasamoto-Nasu-2025}, in which the spin correlation is evaluated after expressing the spin bilinear in terms of the bond operators used in the mean-field Ansatz.
The conventional Wick prescription and its distinction from the bond-operator-based evaluation used here are summarized in Appendix~\ref{app:correlation-prescriptions}.
Starting from the identity
\begin{align}
    \label{eq:spin-correlator-bond-operator-identity}
    S_i^\alpha S_j^{\alpha^{\prime}}
    =
    \sum_{p,q}
    A_{pq}^{\alpha\alpha^{\prime}}
    \normord{\mathcal{Q}_{ij}^{p\dagger}\mathcal{Q}_{ij}^{q}},
\end{align}
we approximate the imaginary-time spin correlation as
\begin{align}
    \label{eq:bond-operator-spin-correlator}
    \chi_{ij}^{\alpha\alpha^{\prime}}(\tau)
    \simeq
    \sum_{p,q}
    A_{pq}^{\alpha\alpha^{\prime}}
    \overline{\mathcal{F}}_{ij}^{p}(\tau)
    \mathcal{F}_{ij}^{q}(\tau).
\end{align}
Here, $\mathcal{F}_{ij}^{q}(\tau)$ and
$\overline{\mathcal{F}}_{ij}^{p}(\tau)$ are the imaginary-time
two-point contractions associated with $\mathcal{Q}_{ij}^{q}$ and
$\mathcal{Q}_{ij}^{p\dagger}$, respectively; their explicit Nambu-space
forms are given in Eqs.~\eqref{eq:bond-operator-contractions-Nambu} and
\eqref{eq:bond-operator-contractions-Nambu-conjugate} below.
To treat hopping- and pairing-type channels in a common notation, we use
the site-local Nambu vector $B_i$ introduced in
Eq.~\eqref{eq:site-Nambu-vector}.
In the same ordering as $\bm{\mathcal Q}_{ij}$, the corresponding
$2\times2$ spin-space matrices are collected in the vector
\begin{align}
    \label{eq:bond-operator-spin-vertices}
    \bm{X}
    =
    \left(
    i\sigma^{y},
    \sigma^{0},
    \sigma^{x},
    \sigma^{y},
    \sigma^{z},
    i\sigma^{y}\sigma^{x},
    i\sigma^{y}\sigma^{y},
    i\sigma^{y}\sigma^{z}
    \right)^{T}.
\end{align}
Thus, $p=1,\ldots,8$ labels the corresponding components of
$\bm{\mathcal Q}_{ij}$ and $\bm X$ in this common ordering.
The associated $4\times4$ Nambu-space vertex is
\begin{align}
    \label{eq:Nambu-bond-vertices}
    \widetilde{X}^{p}
    =
    \begin{cases}
    \begin{pmatrix}
        X^{p} & 0\\
        0 & 0
    \end{pmatrix},
    & p=2,3,4,5,\\[6pt]
    \begin{pmatrix}
        0 & 0\\
        X^{p} & 0
    \end{pmatrix},
    & p=1,6,7,8.
    \end{cases}
\end{align}
All components of $\bm{\mathcal Q}_{ij}$ can then be written in the
common form
\begin{align}
    \label{eq:Nambu-bond-operator}
    \mathcal{Q}_{ij}^{p}
    =
    \frac{1}{2}B_i^{\dagger}\widetilde{X}^{p}B_j.
\end{align}
The distinction between hopping and pairing is encoded in the nonzero
particle--hole block of $\widetilde{X}^{p}$.
The contractions entering
Eq.~\eqref{eq:bond-operator-spin-correlator} are compactly written as
\begin{align}
    \mathcal{F}_{ij}^{p}(\tau)
    &=
    \frac{1}{2}
    \left\langle
    T_{\tau}\overline{B}_{i}(\tau)\widetilde{X}^{p}B_j(0)
    \right\rangle,
    \label{eq:bond-operator-contractions-Nambu}\\
    \overline{\mathcal{F}}_{ij}^{p}(\tau)
    &=
    \frac{1}{2}
    \left\langle
    T_{\tau}\overline{B}_{j}(0)\widetilde{X}^{p\dagger}B_i(\tau)
    \right\rangle.
    \label{eq:bond-operator-contractions-Nambu-conjugate}
\end{align}
At $\tau=0^{-}$, these contractions reduce to the same bond channels that
enter the mean-field Hamiltonian, which is the sense in which this
evaluation of the spin correlation is consistent with the SBMFT Ansatz.
This channel consistency provides the basis for the spin-correlation
prescription employed here; its relation to the direct Wick evaluation of
the Gaussian saddle is discussed in
Appendix~\ref{app:correlation-prescriptions}.
For the dynamical response, these contractions are evaluated by Fourier transforming the Nambu
operators, applying the Bogoliubov transformation in
Eq.~\eqref{eq:Bogoliubov-transformation}, and using the Bose distribution
in Eq.~\eqref{eq:Bose-distribution}.
Substituting them into Eq.~\eqref{eq:bond-operator-spin-correlator} and
performing the spatial and imaginary-time Fourier transform in
Eq.~\eqref{eq:Matsubara-spin-susceptibility} yield
$\chi^{\alpha\alpha^{\prime}}(\bm q,i\omega_n)$, with an internal bosonic
Matsubara-frequency sum arising from the product of the two contractions.
The analytic continuation $i\omega_n\rightarrow\omega+i\delta$ then gives
the retarded response, from which
$S^{\alpha\alpha^{\prime}}(\bm q,\omega)$ follows through
Eq.~\eqref{eq:DSSF-from-Matsubara-susceptibility}.
\section{Results}
\label{sec:results}

In this section, we present our numerical results for the $S=1$ extended Kitaev model.
We first specify the mean-field Ansatz used in the calculation in Sec.~\ref{sec:Mean-field ansatz}.
We then show the zero-temperature dynamical spin structure factor $S(\bm{q},\omega)$, defined in Eq.~\eqref{eq:definition-dynamical-spin-structure-factor}, for the pure $S=1$ Kitaev model within SBMFT in Sec.~\ref{sec:SBMFT-pure-dynamics}.
Finally, in Sec.~\ref{sec:Gamma-Gammap-dynamics}, we present the ground-state spin dynamics of the $S=1$ Kitaev-$\Gamma$-$\Gamma^{\prime}$ model obtained from the $\pi/2$-flux Ansatz.
The auxiliary spectra obtained after retaining only the on-site and nearest-neighbor spin correlations are collected in Appendix~\ref{app:short-range-dynamics}.
In all cases, the spin structure factors are presented as the trace over the three spin components,
\begin{align}
    \label{eq:trace spin structure factor}
    S(\bm{q},\omega)
    &=\frac{1}{3}\sum_{\alpha=x,y,z}S^{\alpha\alpha}(\bm{q},\omega),\\
    S(\bm{q})
    &=\frac{1}{3}\sum_{\alpha=x,y,z}S^{\alpha\alpha}(\bm{q}).
\end{align}
All spectra shown in the Results and in the corresponding Appendices are evaluated in the zero-temperature limit, $\beta\to\infty$.
In this limit, $n_{\rm B}(\varepsilon)=0$ for positive-energy spinon modes.
Unless otherwise noted, the self-consistent SBMFT calculations and the corresponding spectra are evaluated on a $50\times50$ momentum mesh in the Brillouin zone of the eight-sublattice mean-field unit cell.
This convention is used for both the $0$-flux and $\pi/2$-flux Ans\"atze, so that the comparison is made with the same Brillouin-zone folding.
Equivalently, the calculation corresponds to $N=8\times50\times50=20000$ sites in a real-space periodic system.
For the finite broadening parameter in Eq.~\eqref{eq:DSSF-from-Matsubara-susceptibility}, we set $\delta/|K|=0.01$.

\subsection{Mean-field Ansatz}
\label{sec:Mean-field ansatz}

\begin{figure}[t]
  \centering
      \includegraphics[width=\columnwidth,clip]{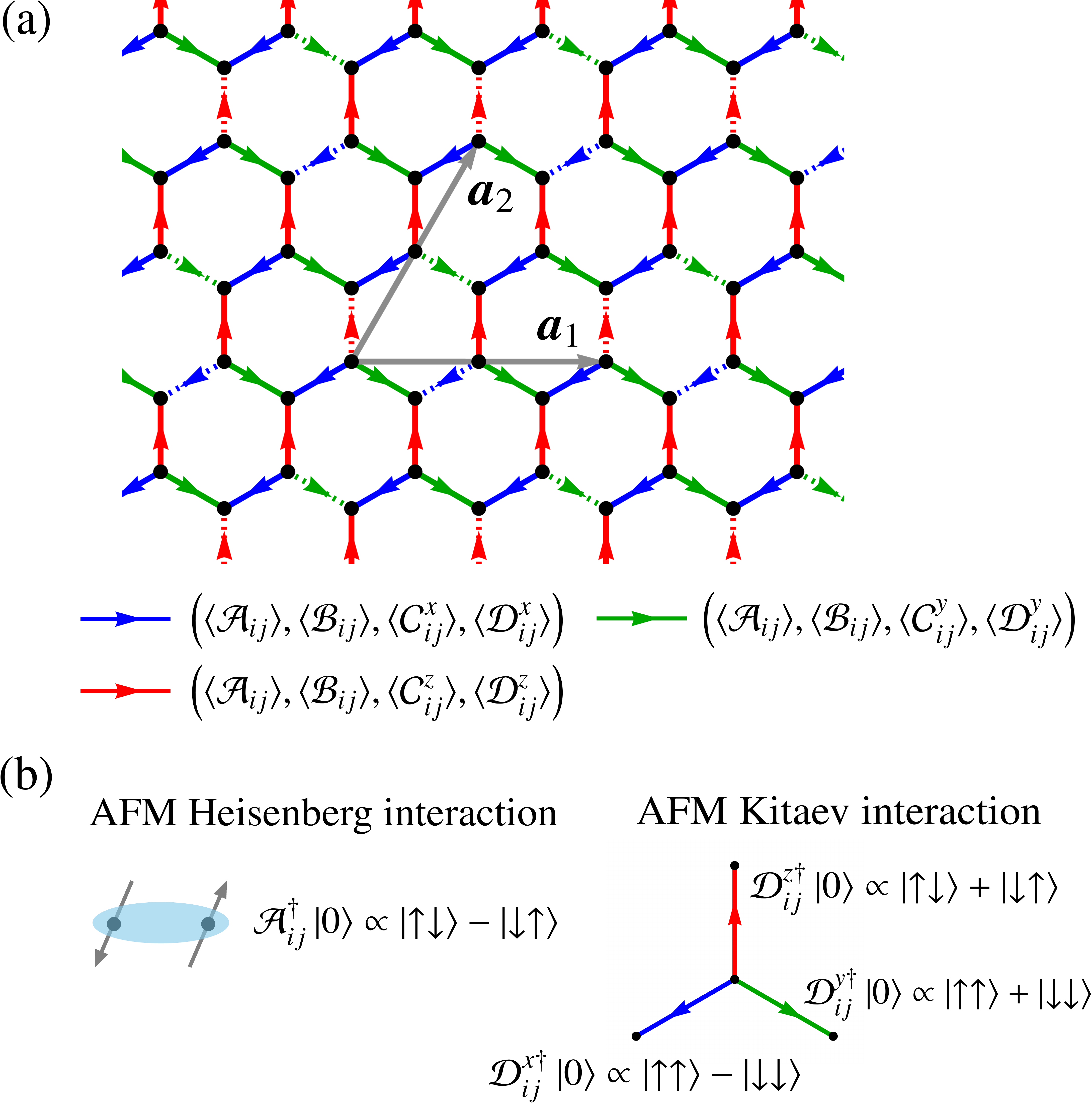}
      \caption{
        (a)
        Mean-field Ansatz for the $S=1$ pure AFM Kitaev model considered in this work. The honeycomb lattice on which the mean fields are defined has an eight-sublattice unit cell, whose primitive translation vectors are indicated by the gray arrows $\bm{a}_{1}$ and $\bm{a}_{2}$. The mean-field parameters defined on the $x$~(blue), $y$~(green), $z$~(red) bonds are summarized in the legend. The direction of each arrow specifies the orientation of the bond operator from site $i$ to $j$. When considering the $\pi/2$-flux state, an additional phase factor $i$ is multiplied to the mean fields on the dotted bonds. 
        (b)
        Channel interpretation of $\mathcal{A}_{ij}$ and $\mathcal{D}_{ij}^{\gamma}$.
        The former represents SU(2)-singlet pairing, whereas the latter represents the bond-dependent $m_\gamma=0$ member of the triplet.
        In the spin-$1/2$ bond picture, the antiferromagnetic Ising ground-state subspace contains both channels.
        Adopting $\mathcal{D}_{ij}^{\gamma}$ as the primary Kitaev mean field is therefore a mean-field prescription.
      }
      \label{fig:mean_field_ansatz}
\end{figure}

Before presenting the structure factors, we describe the mean-field Ansatz used in this work.
The Ansatz is constructed to retain the bond-selective character of the Kitaev interaction while allowing both time-reversal-symmetric and time-reversal-breaking patterns of Schwinger-boson mean fields.
The bond-channel intuition is summarized schematically in Fig.~\ref{fig:mean_field_ansatz}(b).
For an antiferromagnetic Heisenberg exchange, the spin-singlet pairing operator $\mathcal{A}_{ij}$ gives the natural valence-bond picture: because the interaction is SU(2) symmetric, exchange energy is gained through spin-singlet resonance on the bond.
For an antiferromagnetic Kitaev interaction, the local energy gain has a different character.
Consider a $\gamma$ bond connecting site $i$ and its nearest neighbor $i+\delta_{\gamma}$, where $\delta_{\gamma}$ denotes the nearest-neighbor vector associated with the $\gamma$ bond.
The Ising interaction on this bond selects only the $\gamma$ spin component and favors a bond state whose total spin component along the selected axis vanishes, $S_{i}^{\gamma}+S_{i+\delta_{\gamma}}^{\gamma}=0$, in the spin-$1/2$ bond picture.
For $S=1/2$, this condition defines the subspace spanned by $\ket{\uparrow_\gamma\downarrow_\gamma}$ and $\ket{\downarrow_\gamma\uparrow_\gamma}$.
This subspace contains both the SU(2) singlet and the $m_\gamma=0$ member of the triplet.
The local antiferromagnetic Ising interaction alone therefore does not select $\mathcal D_{ij}^{\gamma}$ over $\mathcal A_{ij}$.
Adopting $\mathcal D_{ij}^{\gamma}$ as the primary mean field together with the upper representation in Eq.~\eqref{eq:Heisenberg-interaction-Schwinger-boson-ver2} therefore constitutes a mean-field prescription rather than a consequence of the local Ising energetics alone.
This construction differs from the mixed singlet--triplet hopping-and-pairing decomposition of Ref.~\cite{Ralko-Merino-2024}, in which the relative channel weights were fixed by requiring the three Ising components to reproduce the Heisenberg exchange and its large-$S$ Luttinger--Tisza energy.
The bond-operator basis extends to arbitrary $S$ through the constraint $n_i=2S$.
At the saddle-point level, this constraint is imposed only on average, fixing the average boson number and hence the spin magnitude.

The spatial pattern of the mean fields is shown in Fig.~\ref{fig:mean_field_ansatz}(a).
The enlarged unit cell is chosen so as to accommodate both the four-sublattice transformation connecting the AFM and FM pure Kitaev models and the phase pattern of the chiral Ansatz.
The arrows in the figure define the orientation of the bond operators from $i$ to $j$, and symmetry-inequivalent bonds are allowed to carry independent mean-field amplitudes.
Following the triplet Wilson-loop construction of Ref.~\cite{Ralko-Merino-2024}, we label the flux sector in the triplet-pairing channel by a gauge-invariant phase on an oriented elementary hexagon $i\to j\to k\to l\to m\to n\to i$, whose bond types are taken to be $x,y,z,x,y,z$ in this order.
In the present bond-operator convention, the triplet Wilson-loop product is
\begin{align}
    W_{t}
    &=
    \langle \mathcal{D}^{x}_{ij}\rangle
    \langle \mathcal{D}^{y}_{jk}\rangle^{*}
    \langle \mathcal{D}^{z}_{kl}\rangle
    \langle \mathcal{D}^{x}_{lm}\rangle^{*}
    \langle \mathcal{D}^{y}_{mn}\rangle
    \langle \mathcal{D}^{z}_{ni}\rangle^{*},
    \label{eq:triplet-Wilson-loop}
\end{align}
and the corresponding flux $\phi_{t}$ is defined by
\begin{align}
    e^{i\phi_{t}}=\frac{W_{t}}{|W_{t}|}.
    \label{eq:triplet-flux}
\end{align}
The phase $\phi_t$ is defined modulo $2\pi$ and is invariant under local U(1) gauge transformations of the Schwinger bosons.
The $0$-flux Ansatz used below has $\phi_t=0$.
For the $\pi/2$-flux Ansatz, an additional phase factor $i$ is assigned to the dotted bonds in Fig.~\ref{fig:mean_field_ansatz}(a), giving $\phi_t=\pi/2$ in the above convention.
The mean-field parameters are optimized within this constrained $\pi/2$-flux manifold.
Since time reversal complex conjugates the mean fields and sends $\phi_t\to-\phi_t$, the $\pi/2$-flux Ansatz breaks time-reversal symmetry at the mean-field level.
We do not consider the triplet $\pi$-flux sector in the present work, because previous SBMFT calculations found that this sector does not remain a stable gapped spin-liquid solution at the physical spin values relevant here; it instead enters a boson-condensed, magnetically ordered regime already at small spin length, of order $S\simeq0.25$~\cite{Ralko-Merino-2024}.
Thus, the $0$-flux and $\pi/2$-flux Ans\"atze compared below should be viewed as different flux patterns built from the same Kitaev-motivated, bond-selective bosonic channel.
The singlet and triplet $\pi/2$-flux solutions are degenerate at the pure Kitaev point within the present mean-field decomposition, as in Ref.~\cite{Ralko-Merino-2024}. In a singlet-only Ansatz, the spin-dependent hopping and triplet-pairing amplitudes satisfy $\langle\mathcal C_{ij}^{\mu}\rangle=\langle\mathcal D_{ij}^{\mu}\rangle=0$. Equation~\eqref{eq:off-diagonal-interaction-Schwinger-boson} then gives no mean-field contribution from either $\Gamma$ or $\Gamma^{\prime}$. At fixed $K$, the self-consistent fields, mean-field energy, and spinon gap therefore retain their pure-Kitaev values within this restricted Ansatz. In contrast, the triplet $\pi/2$-flux solution remains gapped and lowers its energy at $\Gamma/|K|=\Gamma^{\prime}/|K|=\pm0.025$ [see Fig.~\ref{fig:ED_Energy}]. It therefore has lower energy than the restricted singlet solution at these couplings, motivating its use for the spectra at nonzero $\Gamma=\Gamma^{\prime}$.

\subsection{SBMFT dynamics in the pure Kitaev limit}
\label{sec:SBMFT-pure-dynamics}

\begin{figure*}[t]
  \centering
      \includegraphics[width=2\columnwidth,clip]{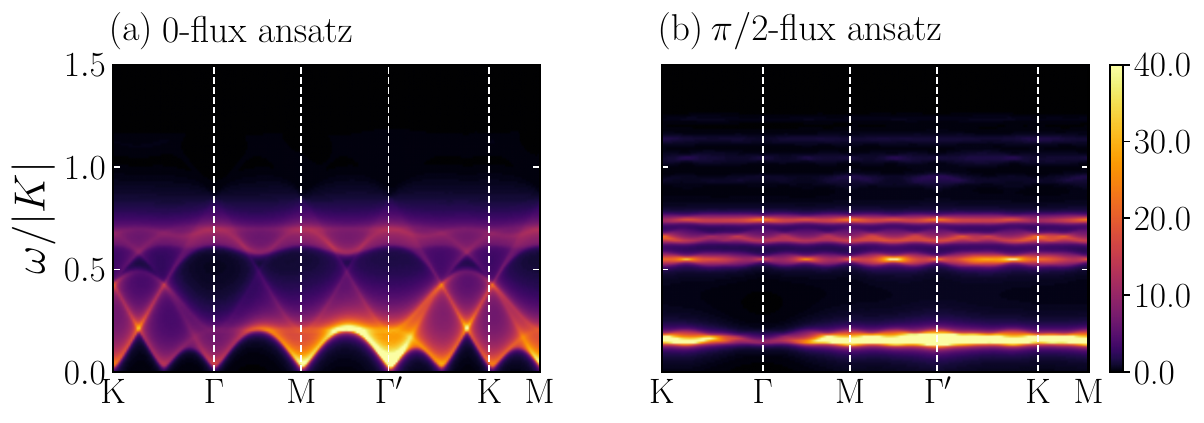}
      \caption{
        Zero-temperature dynamical spin structure factor $S(\bm{q},\omega)$ of the pure antiferromagnetic $S=1$ Kitaev model calculated directly within SBMFT.
        Panels (a) and (b) show the $0$-flux and $\pi/2$-flux Ans\"atze, respectively.
        The spectra are plotted along the path connecting the high-symmetry points indicated in Fig.~\ref{fig:honeycomb_lattice}.
        The color scale, common to both panels, indicates the intensity of $S(\bm{q},\omega)$, and the broadening parameter is set to $\delta/|K|=0.01$.
      }
      \label{fig:DSSF_AFM_SBMFT}
\end{figure*}

We first present the zero-temperature dynamical spin structure factor of the pure $S=1$ Kitaev model obtained directly from SBMFT.
A comparison of the same triplet $0$- and $\pi/2$-flux Ans\"atze at the pure Kitaev point, evaluated using the bond-operator correlation prescription, was previously reported in Ref.~\cite{Sasamoto-Nasu-2025}.
Here, we revisit this comparison to relate the dynamical spectra to the real-space spin correlations and to establish the reference point for the finite-$\Gamma,\Gamma^{\prime}$ analysis below.
Since the present material motivation concerns antiferromagnetic Kitaev exchange, the main text focuses on $K>0$ and compares the two flux patterns on the same footing.
The corresponding FM spectra, obtained after implementing the four-sublattice mapping in the SBMFT bond-operator convention, are shown in Appendix~\ref{app:four-sublattice-transformation} [see Fig.~\ref{fig:DSSF_FM_SBMFT}].
Figure~\ref{fig:DSSF_AFM_SBMFT} compares the $0$-flux Ansatz and the $\pi/2$-flux Ansatz at $T=0$.
For the $0$-flux Ansatz, the spectrum has pronounced momentum dependence at low energy.
Strong spectral weight appears around the $\mathrm{M}$--$\Gamma^{\prime}$ part of the path, together with additional low-energy intensity near the end of the path and a broader higher-energy component around $\omega/|K|\simeq0.6$--$0.8$.
The low-energy response is not a single horizontal feature; rather, it consists of several momentum-dependent ridges whose intensities vary along the path.
By contrast, Fig.~\ref{fig:DSSF_AFM_SBMFT}(b) shows that the $\pi/2$-flux Ansatz gives a visibly flatter response.
The strongest low-energy intensity is distributed around the $\mathrm{M}$--$\Gamma^{\prime}$--$\mathrm{K}$ portion of the path, and a broad higher-energy feature remains near $\omega/|K|\simeq0.6$.
Compared with the $0$-flux result, the low-energy weight forms a more weakly momentum-dependent, nearly horizontal feature over a wider portion of the path.
Analytical expressions for the pure-Kitaev spinon gaps and a complementary comparison of the spinon dispersions and static structure factors are given in Appendix~\ref{app:a}.

\subsection{Off-diagonal-exchange dynamics}
\label{sec:Gamma-Gammap-dynamics}

We next examine the ground-state spin dynamics under weak same-sign off-diagonal exchanges along $\Gamma^{\prime}=\Gamma$. We use the $\pi/2$-flux Ansatz, which remains gapped at $\Gamma/|K|=\Gamma^{\prime}/|K|=\pm0.025$. Checks on momentum meshes up to $200\times200$ confirm that the $0$-flux branch also has a narrow finite-gap region around the pure Kitaev point, but develops unstable spinon modes at much smaller exchange magnitudes.

\begin{figure*}[t]
  \centering
  \includegraphics[width=2\columnwidth]{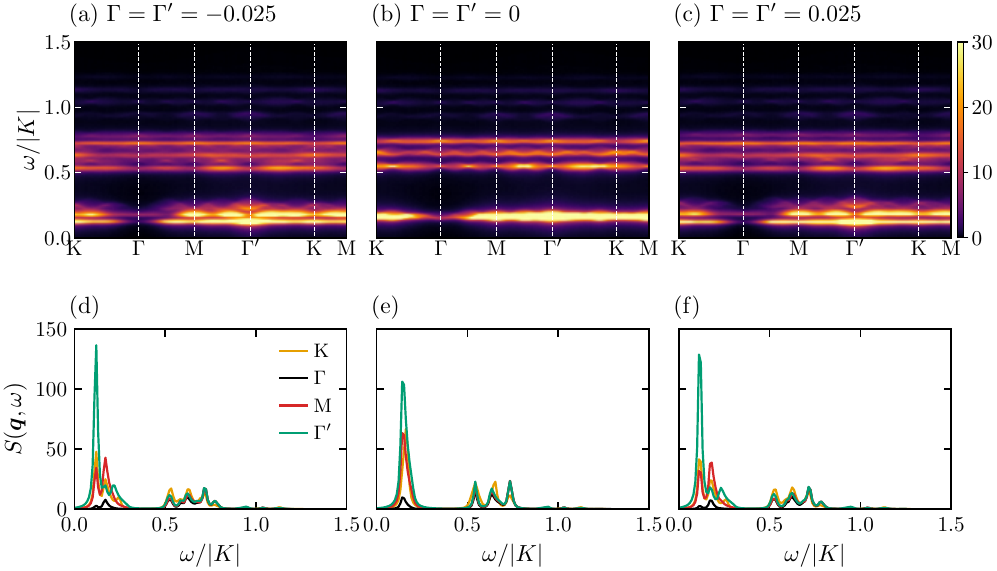}
  \caption{
    Ground-state spin dynamics of the $S=1$ Kitaev-$\Gamma$-$\Gamma^{\prime}$ model calculated within SBMFT using the $\pi/2$-flux Ansatz for $K>0$.
    The upper row shows the dynamical spin structure factor $S(\bm{q},\omega)$ along the momentum path for (a) $\Gamma=\Gamma^{\prime}=-0.025$, (b) $0$, and (c) $0.025$, in units of $|K|$.
    Panels (d)--(f) show the corresponding energy dependence at the representative high-symmetry momenta $\mathrm{K}$, $\Gamma$, $\mathrm{M}$, and $\Gamma^{\prime}$.
    A common color scale is used in panels (a)--(c), and a common vertical intensity scale is used in panels (d)--(f).
    The broadening parameter is $\delta/|K|=0.01$.
  }
  \label{fig:Gamma_DSSF}
\end{figure*}

Figure~\ref{fig:Gamma_DSSF} compares three points on the same-sign line $\Gamma^{\prime}=\Gamma$, from $\Gamma/|K|=-0.025$ to $0.025$.
Panels~\ref{fig:Gamma_DSSF}(a)--\ref{fig:Gamma_DSSF}(c) show the momentum-resolved dynamical spin structure factor, while panels~\ref{fig:Gamma_DSSF}(d)--\ref{fig:Gamma_DSSF}(f) show cuts at the representative high-symmetry momenta.
Starting from the pure Kitaev result in panels~\ref{fig:Gamma_DSSF}(b) and \ref{fig:Gamma_DSSF}(e), the finite off-diagonal exchanges visibly redistribute the low-energy spectral weight.
For both signs, the strongest low-energy response remains at $\Gamma^{\prime}$, and the corresponding peak occurs at or below the dominant low-energy peaks at $\mathrm{K}$ and $\mathrm{M}$.
All three parameter sets also exhibit weaker finite-energy structures around $\omega/|K|\simeq0.5$--$0.8$.
Changing the common sign of $\Gamma$ and $\Gamma^{\prime}$ modifies the detailed low-energy line shape and momentum dependence, while the dominant $\Gamma^{\prime}$ response and the finite-energy spectral weight remain visible throughout the displayed interval.

\section{Discussion}
\label{sec:Discussion}

\begin{figure*}[t]
  \centering
  \includegraphics[width=2\columnwidth]{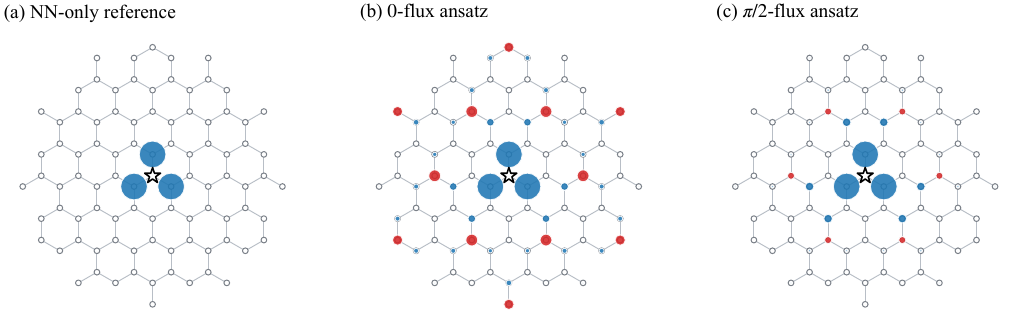}
  \caption{
    Real-space distribution of the equal-time spin correlation
    $C_{ij}=\langle\bm{S}_{i}\cdot\bm{S}_{j}\rangle$ in the pure antiferromagnetic $S=1$ Kitaev limit.
    The star denotes the reference site $i$.
    Blue and red circles indicate negative and positive values of $C_{ij}$, respectively.
    In each panel, the symbol area is proportional to $|C_{ij}|/|C_{\mathrm{NN}}|$, where $|C_{\mathrm{NN}}|$ is the largest magnitude among the three nearest-neighbor correlations in that panel.
    The nearest-neighbor correlations are thereby normalized to unity and shown with the same symbol area in all three panels.
    Panel (a) shows the nearest-neighbor reference. Exact Kitaev locality fixes the absence of longer-distance correlations~\cite{Baskaran-Sen-Shankar-2008}, whereas the retained value $C_{ij}=-0.4330$ is estimated from exact diagonalization on an $N=24$ cluster~\cite{Koga-2018}.
    Panels (b) and (c) show the SBMFT results for the $0$-flux and $\pi/2$-flux Ans\"atze, respectively.
  }
  \label{fig:spin_correlator_realspace}
\end{figure*}

Figure~\ref{fig:spin_correlator_realspace} compares the spatial distribution of the equal-time spin correlation $C_{ij}=\langle\bm S_i\cdot\bm S_j\rangle$ around a fixed reference site for (a) the nearest-neighbor Kitaev reference, (b) the $0$-flux SBMFT Ansatz, and (c) the $\pi/2$-flux SBMFT Ansatz.
The color of each circle indicates the sign of $C_{ij}$, while its area represents $|C_{ij}|$ normalized by the largest nearest-neighbor magnitude in the corresponding panel.
This representation allows the spatial decay of the correlations to be compared relative to the nearest-neighbor scale.
The nearest-neighbor reference in Fig.~\ref{fig:spin_correlator_realspace}(a) combines the nearest-neighbor correlation obtained by exact diagonalization on an $N=24$ cluster~\cite{Koga-2018} with the exact locality of the pure Kitaev model~\cite{Baskaran-Sen-Shankar-2008}: apart from the on-site term, only the three nearest-neighbor Kitaev-bond correlations connected to the reference site are retained.
Both SBMFT Ans\"atze reproduce the dominant negative nearest-neighbor correlations, but their longer-distance components are different.
In the $0$-flux Ansatz, visible positive and negative correlations persist at multiple distances from the reference site, whereas in the $\pi/2$-flux Ansatz the longer-distance correlations are strongly reduced relative to the nearest-neighbor scale.
This comparison shows that, within the present SBMFT calculation, the $\pi/2$-flux Ansatz incorporates the short-range character of the Kitaev spin correlations more effectively than the $0$-flux Ansatz.

This observation should not be read as a statement that the $0$-flux Ansatz is intrinsically inappropriate for the pure $S=1$ Kitaev problem.
An ordinary SBMFT Ansatz does not impose, as an exact constraint, the vanishing of spin correlations beyond nearest-neighbor Kitaev bonds.
The additional longer-distance correlations in Fig.~\ref{fig:spin_correlator_realspace}(b) should therefore be viewed as a limitation of this particular mean-field realization of the $0$-flux Ansatz, rather than as a direct exclusion of a time-reversal-symmetric description of the pure Kitaev limit.
This point is important because the higher-spin Kitaev problem still possesses exact local conserved quantities, and the corresponding gauge structure itself does not require time-reversal-symmetry breaking~\cite{Baskaran-Sen-Shankar-2008,Ma-2023}.
Short-ranged equal-time correlations constrain the frequency-integrated structure factor, $\int d\omega\,S(\bm q,\omega)$, but do not uniquely determine its frequency-resolved momentum dependence.
The auxiliary truncation in Appendix~\ref{app:short-range-dynamics} shows that removing the longer-range mean-field correlations also makes the $0$-flux response nearly flat.
Consequently, spectral flatness alone provides a consistency check for Kitaev-like locality within the present prescription, but does not identify the Schwinger-boson flux sector or establish time-reversal breaking in the microscopic ground state.

To place the gapped, short-ranged character of the present $\pi/2$-flux saddle in a broader theoretical context, we next relate it to two complementary parton descriptions of higher-spin Kitaev physics.
As reviewed in Sec.~\ref{introduction}, the general-$S$ gauge construction of Ref.~\cite{Ma-2023} places $S=1$ in the integer-spin branch with a bosonic gauge charge, for which a gapped deconfined $\mathbb Z_2$ phase is allowed.
The multilayer Kitaev model provides a complementary fermionic-parton realization of the same even--odd structure~\cite{Merino-Ralko-2025}.
Within the Abrikosov-fermion description, the liquid saddle points are found to be gapped for an even number of layers and gapless for an odd number of layers.
For a bilayer, a gapped saddle appears around $J_{\perp}/|K|\simeq 2.5$ and may be viewed as two Kitaev layers whose $\mathbb Z_2$ bond fields have opposite signs.
The same saddle develops interlayer Cooper pairing of the fermionic partons, producing effectively bosonic composite degrees of freedom and an effective local spin larger than $1/2$.
The alternating bond fields break inversion between the two layers and gap the Majorana cones of the individual Kitaev layers.
In the flux convention of Ref.~\cite{Merino-Ralko-2025}, this pattern is described by an in-plane $0$ flux and an interlayer $\pi$ flux.
In the limit of strong ferromagnetic interlayer coupling, $n$ layers project onto a spin-$S=n/2$ Kitaev model.
The explicit relation between Abrikosov and Majorana bond variables leads to a multilayer PSG characterized by the alternating bond signs.
The agreement between the Gutzwiller-projected and exact strong-coupling energies provides additional support for this connection.

\begin{figure}[t]
  \centering
  \includegraphics[width=\columnwidth]{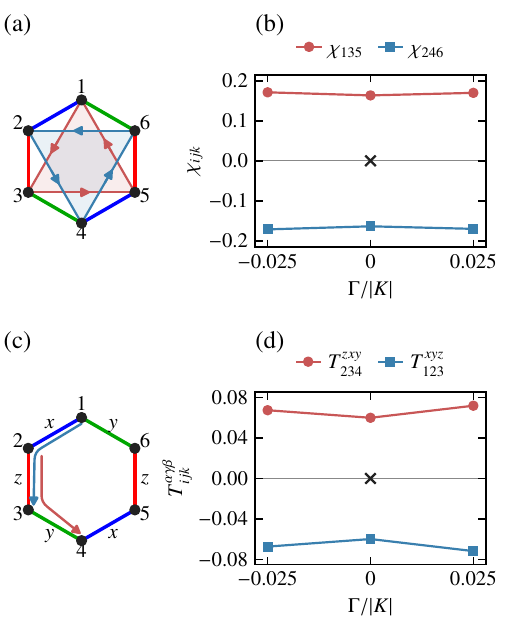}
  \caption{Three-spin correlations of the triplet $+\pi/2$-flux saddle for $K>0$ and $\Gamma^{\prime}=\Gamma$. (a,b) Counterclockwise alternating-site triangles and scalar chiralities $\chi_{135},\chi_{246}$. (c,d) Consecutive-site paths and Kitaev-type components $T_{123}^{xyz},T_{234}^{zxy}$. Bond colors follow Fig.~\ref{fig:honeycomb_lattice}(a). Arrows indicate site ordering. Crosses show the pure $0$-flux reference at $\Gamma=\Gamma^{\prime}=0$.}
  \label{fig:three-spin-chirality}
\end{figure}

The strong-coupling relation and even--odd gap structure place the bilayer saddle in the effectively bosonic integer-spin branch discussed above.
Its Kitaev-like structure motivates predominantly short-ranged spin correlations and a gapped, weakly dispersive dynamical response, consistent with numerical studies of the $S=1$ Kitaev model~\cite{Lee-Kawashima-Kim-2020,Dong-Sheng-2020,Khait-Stavropoulos-2021,Chen-2022}.
These properties provide a useful point of comparison with the present gapped $\pi/2$-flux SBMFT saddle: Fig.~\ref{fig:spin_correlator_realspace} shows that its correlations are shorter ranged than those of the $0$-flux saddle and closer to the nearest-neighbor Kitaev reference, while Fig.~\ref{fig:DSSF_AFM_SBMFT} shows the corresponding flatter dynamical response.
Nevertheless, the two constructions employ different parton representations and realize different broken symmetries: the present $\pi/2$-flux Ansatz breaks time reversal, whereas the bilayer Abrikosov-fermion saddle breaks interlayer inversion.
Because their flux labels are representation dependent, these similarities do not establish an equivalence between the two mean-field constructions.
Within the present Schwinger-boson description, the nonzero gauge-invariant Wilson phase identifies the $\pi/2$-flux Ansatz as time-reversal breaking~\cite{Ralko-Merino-2024}: time reversal maps the $\phi_t=+\pi/2$ saddle onto the distinct $\phi_t=-\pi/2$ saddle.
To test this property directly with spin operators, we evaluate the equal-time three-spin correlations
\begin{align}
    \chi_{ijk}&=\bigl\langle\bm S_i\cdot(\bm S_j\times\bm S_k)\bigr\rangle,
    \label{eq:scalar-three-spin-chirality}\\
    T_{ijk}^{\alpha\gamma\beta}
    &=\bigl\langle S_i^\alpha S_j^\gamma S_k^\beta\bigr\rangle.
    \label{eq:Kitaev-three-spin-correlation}
\end{align}
The site ordering for $\chi_{ijk}$ is illustrated in Fig.~\ref{fig:three-spin-chirality}(a). In the second expression, $i,j,k$ are consecutive sites on an $\alpha$ bond and a $\beta$ bond [see Fig.~\ref{fig:three-spin-chirality}(c)], and $\gamma$ is the remaining spin component. Both quantities are odd under physical time reversal. We compute them by Wick contraction of the six Schwinger-boson operators in the self-consistently converged, unprojected Gaussian ground state. This evaluates three-spin expectation values of the saddle itself, rather than extending the bond-operator prescription used for the two-spin dynamical response.

Figure~\ref{fig:three-spin-chirality} shows finite staggered three-spin correlations at $\Gamma/|K|=\Gamma^{\prime}/|K|=0,\pm0.025$. At the pure Kitaev point, $\chi_{135}\simeq0.16339$ and $\chi_{246}=-\chi_{135}$ for the two counterclockwise triangles. The consecutive-site components also alternate in sign, with $\langle S_1^x S_2^y S_3^z\rangle\simeq-0.059827$ and $\langle S_2^z S_3^x S_4^y\rangle\simeq0.059827$. The same sign pattern holds on all four hexagons of the mean-field unit cell at each sampled coupling. The sum of the two scalar chiralities and the sum of the six consecutive-site components vanish, so these results describe staggered, not uniform, chirality. The pure $0$-flux reference gives zero for both observables.
The saddle preserves $PT$ up to a Schwinger-boson gauge transformation, where $P$ is inversion about a hexagon center and $T$ is physical time reversal. In Fig.~\ref{fig:three-spin-chirality}(a), $P$ exchanges the two counterclockwise triangles and $T$ reverses their chirality, so $PT$ relates $\chi_{135}$ to $-\chi_{246}$. Time reversal alone reverses the three-spin correlations and maps the $+\pi/2$ saddle to its $-\pi/2$ partner of equal energy. The nonzero staggered three-spin correlations therefore support the conclusion that the triplet $\pi/2$-flux state breaks time-reversal symmetry within SBMFT.

For finite same-sign exchanges $\Gamma=\Gamma^{\prime}$ at the parameter points shown in Fig.~\ref{fig:Gamma_DSSF}, a gapped solution is obtained within the constrained $\pi/2$-flux manifold.
The $0$-flux branch also admits a much narrower finite-gap continuation, but softens at substantially smaller exchanges, as summarized in Sec.~\ref{sec:Gamma-Gammap-dynamics}.
Within a different SBMFT decomposition, Ref.~\cite{Ralko-Merino-2024} found a chiral bosonic saddle to remain stable over a broad range of spin magnitude.
The spin-liquid regime reported by pf-FRG~\cite{Fukui-Kato-Nasu-Motome-2022} and the dynamics of Ref.~\cite{Sasamoto-Nasu-2025} are likewise consistent with Kitaev-like behavior.

\begin{figure}[t]
  \centering
  \includegraphics[width=\columnwidth]{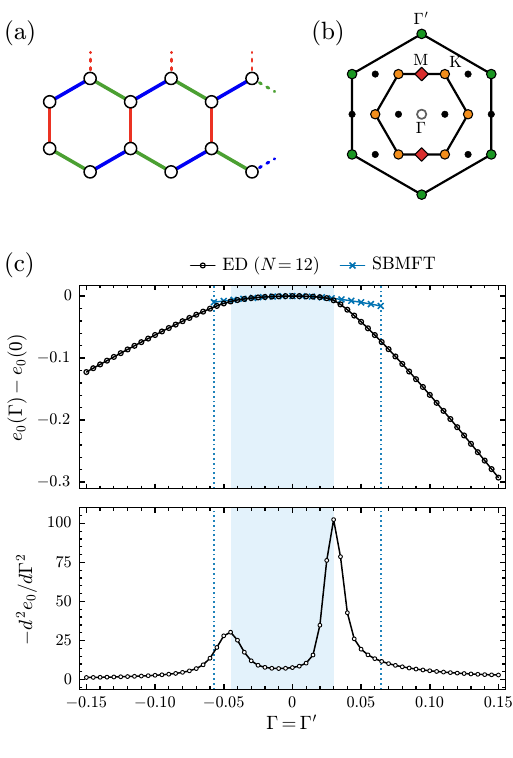}
  \caption{
    Finite-cluster geometry and ground-state energy comparison on the line $\Gamma=\Gamma^{\prime}$.
    (a) Real-space shape of the $N=12$ honeycomb cluster used in the exact-diagonalization calculation.
    The colored bonds indicate the three Kitaev bond directions, and the dashed bonds represent the periodic boundary condition.
    (b) Allowed momenta of the same cluster in the extended Brillouin-zone representation.
    Selected high-symmetry momenta are highlighted.
    (c) Ground-state energy change $e_{0}(\Gamma)-e_{0}(0)$, where $e_{0}=E_{0}/N$ is the ground-state energy per site, and its second derivative along $\Gamma=\Gamma^{\prime}$.
    Black circles denote the exact-diagonalization results for $N=12$, while blue crosses denote the $\pi/2$-flux SBMFT solution in the gapped spin-liquid regime.
    The shaded region denotes the finite-size Kitaev-like window whose boundaries are determined from the dominant peaks in $-d^{2}e_{0}/d\Gamma^{2}$, and the blue dotted lines mark the mean-field spinon-condensation instabilities of the $\pi/2$-flux SBMFT solution.
  }
  \label{fig:ED_Energy}
\end{figure}

The energy comparison in Fig.~\ref{fig:ED_Energy} provides a complementary finite-size benchmark for the same line $\Gamma=\Gamma^{\prime}$.
In the $N=12$ ED result, the ground-state energy changes only weakly within the shaded Kitaev-like window, while it decreases much more rapidly once the system moves into the neighboring ordered regimes. 
The shaded window is bounded by the dominant peaks in $-d^{2}e_{0}/d\Gamma^{2}$ on the negative- and positive-$\Gamma$ sides, which provide a finite-size indication of these crossovers or instabilities.
Within the central window where the SBMFT solution remains gapped, the $\pi/2$-flux result reproduces the weak energy variation caused by the off-diagonal exchanges.
To quantify this comparison, we consider the energy change $e_0(\Gamma)-e_0(0)$ separately for each method. At $\Gamma/|K|=\Gamma^{\prime}/|K|=-0.025$, ED and SBMFT give $[e_0(\Gamma)-e_0(0)]/|K|=-0.0022561$ and $-0.0023388$, respectively, a relative deviation of $3.7\%$ from ED. At $+0.025$, the corresponding values are $-0.0034030$ and $-0.0027583$, with a larger deviation of $18.9\%$. Dividing the magnitude of the energy change at $\Gamma=\Gamma^{\prime}>0$ by that at $\Gamma=\Gamma^{\prime}<0$ gives $1.51$ for ED and $1.18$ for SBMFT. Thus, the mean-field result captures the weak energy response particularly closely for $\Gamma=\Gamma^{\prime}<0$ but underestimates this sign asymmetry.
The blue dotted lines in Fig.~\ref{fig:ED_Energy} denote the $\pi/2$-flux SBMFT spinon-condensation instabilities, namely the mean-field estimates for the boundary of the gapped saddle.
The $\pi/2$-flux SBMFT instability points lie outside the shaded $N=12$ ED window.
This indicates that the constrained mean-field saddle remains gapped over a broader parameter interval, particularly for $\Gamma>0$.
The shaded ED boundaries are finite-size pseudocritical scales identified from the peaks in $-d^{2}e_{0}/d\Gamma^{2}$, and the $N=12$ calculation alone does not determine their thermodynamic limits. Nevertheless, DMRG calculations for the spin-$1$ Kitaev-$\Gamma$ model with $\Gamma^{\prime}=0$ found that the antiferromagnetic Kitaev liquid is destabilized already by a small positive $\Gamma/|K|\simeq0.028$~\cite{Luo-Zhao-Li-Wang-2024}. Although that parameter line differs from the present $\Gamma^{\prime}=\Gamma$ line, it supports the possibility that the narrow positive-$\Gamma$ window found by ED reflects a genuine instability. The two calculations nevertheless exhibit different sign asymmetries. The ED window extends farther on the negative-$\Gamma$ side, whereas the constrained SBMFT boundary extends only slightly farther on the positive-$\Gamma$ side. This discrepancy may reflect a limitation of the saddle-point description.

\begin{figure*}[t]
  \centering
  \includegraphics[width=\textwidth]{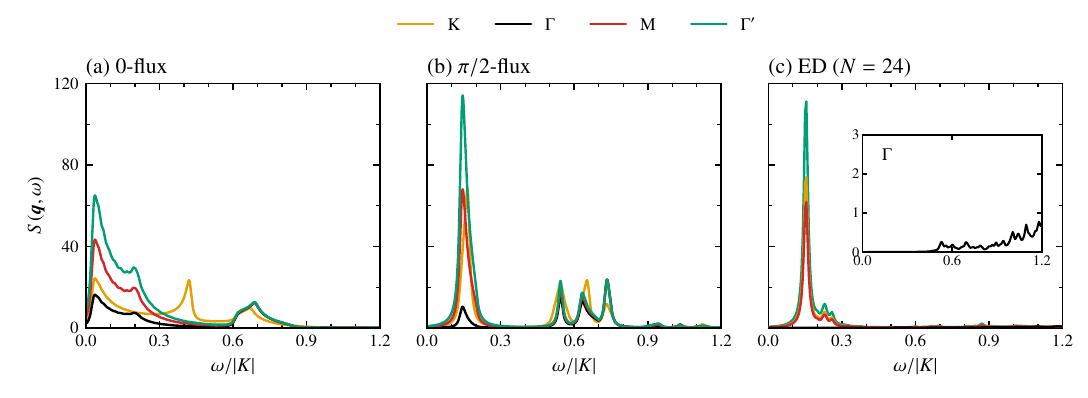}
  \caption{Dynamical spin structure factor of the antiferromagnetic $S=1$ Kitaev model at $\Gamma=\Gamma^{\prime}=0$ for (a) the triplet $0$-flux SBMFT Ansatz, (b) the triplet $\pi/2$-flux SBMFT Ansatz, and (c) ED on a periodic $N=24$ cluster. The curves show $S(\bm{q},\omega)$ at $\mathrm{K}$, $\Gamma$, $\mathrm{M}$, and $\Gamma^{\prime}$ with the same broadening $\eta/|K|=0.01$ and the same main-panel axis scales. In (c), the dominant peaks at $\mathrm{K}$, $\mathrm{M}$, and $\Gamma^{\prime}$ occur at the same frequency. The inset shows the $\Gamma$ spectrum on an expanded intensity scale.}
  \label{fig:ED_Sqw}
\end{figure*}

Figure~\ref{fig:ED_Sqw} compares the two triplet Ans\"atze with $N=24$ ED at the pure Kitaev point. At $\mathrm{K}$, $\mathrm{M}$, and $\Gamma^{\prime}$, both ED and the triplet $\pi/2$-flux Ansatz show sharp, dominant low-energy peaks, at $\omega/|K|\simeq0.155$ in ED and between $0.145$ and $0.164$ for the triplet $\pi/2$-flux Ansatz. By contrast, the $0$-flux spectra peak much closer to zero frequency, at $\omega/|K|\simeq0.037$, and decay more gradually above those peaks. At $\Gamma$, neither Ansatz reproduces the weak ED response in the energy range of the mean-field peaks. The additional spectral weight at higher energies in Fig.~\ref{fig:ED_Sqw}(b) may partly reflect the enforcement of the local boson-number constraint only on average~\cite{Ralko-Merino-2026-Honeycomb}. This comparison supports the triplet $\pi/2$-flux Ansatz as a useful mean-field reference state for the low-energy spin dynamics of the pure Kitaev model.

\section{Summary}
\label{sec:Summary}

We have reexamined the triplet $\phi_t=\pi/2$ chiral bosonic Ansatz introduced in Ref.~\cite{Ralko-Merino-2024} for the $S=1$ Kitaev model and extended the analysis to the Kitaev-$\Gamma$-$\Gamma^{\prime}$ model. Using the Schwinger-boson bond-operator formulation and the spin-correlation prescription of Ref.~\cite{Sasamoto-Nasu-2025}, we calculated the dynamical spin structure factor for both the $0$-flux and $\pi/2$-flux Ans\"atze. In the pure Kitaev model, the $\pi/2$-flux response is flatter than the $0$-flux response, and its real-space spin correlations suppress longer-distance components more strongly and more closely approach the nearest-neighbor Kitaev reference. The $\pi/2$-flux Ansatz also better captures the dominant low-energy peaks of the $N=24$ ED spectra at the pure Kitaev point. Motivated by theoretical proposals for antiferromagnetic Kitaev exchange in spin-$1$ Ni-based systems~\cite{Stavropoulos-2019} and by experimental analyses reporting same-sign off-diagonal exchanges~\cite{Samarakoon-2021}, we further examined weak perturbations on the line $\Gamma^{\prime}=\Gamma$. For the representative perturbations $\Gamma=\Gamma^{\prime}=\pm0.025$, gapped solutions are obtained within the constrained $\pi/2$-flux manifold. On the $N=12$ cluster, ED shows weak ground-state-energy variations within the central Kitaev-like window. The short-ranged spin correlations and characteristic dynamical response make the triplet $\pi/2$-flux Ansatz a useful reference state for studying spin dynamics in the weak Kitaev-$\Gamma$-$\Gamma^{\prime}$ regime. The staggered three-spin correlations provide evidence for time-reversal symmetry breaking in the triplet $\pi/2$-flux mean-field state, supporting its interpretation as a chiral spin-liquid Ansatz.

An important direction for future work is to examine the dynamical response and three-spin correlations of the microscopic model on larger clusters using high-accuracy numerical methods such as DMRG. Finite-size analyses are needed to assess whether the spectral and chiral features found at the mean-field level persist in the thermodynamic limit.
Toward material-specific applications, quantitative comparisons with candidate magnets will require further interactions omitted here, notably Heisenberg exchange and crystal-field-induced single-ion anisotropy~\cite{Stavropoulos-2019,Bradley-2022}. Although these terms can in principle be incorporated into an effective spin-$1$ treatment, an SU($3$) flavor-boson formulation, in which the three local spin states are represented by three bosonic flavors, provides a framework for treating dipolar and quadrupolar fluctuations on equal footing~\cite{Zhang-Wierschem-2013,Muniz-Kato-Batista-2014}. Related SU($3$) approaches have also been used to describe dynamical spectra of spin-$1$ magnets with strong single-ion anisotropy~\cite{Do-2023}, suggesting a route toward material-specific extensions of the present calculation. Further spectroscopic measurements on candidate spin-$1$ Kitaev magnets, combined with these theoretical developments, could test the predicted dynamical signatures and clarify the nature of the underlying quantum states.

\begin{acknowledgments}
  D.S. and J.N. thank Y.~Kamiya for valuable advice and fruitful discussions.
  Parts of the numerical calculations were performed in the supercomputing systems in ISSP, the University of Tokyo.
  This work was supported by Grant-in-Aid for Scientific Research from
  JSPS, KAKENHI Grant Nos.~JP23H01129, JP23H04865, JP24K00563, JP26H00624, JP26H02230.
  D.S. acknowledges support from GP-Spin at Tohoku University.
  A.R. acknowledges financial support from the ANR FlatMoi project (ANR-21-CE30-0029).
  J.M. acknowledges financial support from MICIN/FEDER, Uni\'on Europea, under Grant No.~PID2022-139995NB-I00, and from the Mar\'ia de Maeztu Programme for Units of Excellence in R\&D, Grant No.~CEX2023-001316-M.
\end{acknowledgments}

\appendix

\section{Correlation prescriptions and zeroth-moment sum rule}
\label{app:correlation-prescriptions}

At a fixed quadratic SBMFT saddle, spin correlations may be evaluated either by direct Wick contraction or by the bond-operator prescription used in the main text.
The two prescriptions are generally inequivalent.
They yield markedly different dynamical and equal-time structure factors for the SU(2)-breaking triplet $\phi_t=0$ and $\phi_t=\pi/2$ Ans\"atze, whereas they give essentially identical results for the SU(2)-invariant singlet $\phi_s=\pi$ and $\phi_s=\pi/2$ Ans\"atze.
These explicit comparisons are presented in Ref.~\cite{Sasamoto-Nasu-2025}.
We first summarize their formal distinction and then compare the integrated spectral weight obtained from each prescription with the exact zeroth-moment sum rule.

The conventional prescription applies Wick's theorem directly to the four-boson correlator generated by the spin operators~\cite{Ralko-Merino-2024,Sasamoto-Nasu-2025}.
For a spin-liquid saddle with $\langle\bm S_i\rangle=0$, the disconnected contribution vanishes and
\begin{align}
    \label{eq:Wick-spin-correlator}
    \chi_{ij,\mathrm{W}}^{\alpha\alpha^{\prime}}(\tau)
    &=
    \frac{1}{4}
    \sum_{\mu,\nu,\rho,\lambda}
    \sigma_{\mu\nu}^{\alpha}
    \sigma_{\rho\lambda}^{\alpha^{\prime}}
    \Big[
    \langle T_{\tau}\bar b_{i\mu}(\tau)\bar b_{j\rho}(0)\rangle
    \langle T_{\tau}b_{i\nu}(\tau)b_{j\lambda}(0)\rangle
    \notag\\
    &\qquad+
    \langle T_{\tau}\bar b_{i\mu}(\tau)b_{j\lambda}(0)\rangle
    \langle T_{\tau}b_{i\nu}(\tau)\bar b_{j\rho}(0)\rangle
    \Big].
\end{align}
Because $\mathcal H_{\mathrm{MF}}$ is quadratic, Eq.~\eqref{eq:Wick-spin-correlator} gives the exact Wick evaluation of the spin correlator in the Gaussian ensemble defined by $\mathcal H_{\mathrm{MF}}$.
At a fixed BdG saddle, it includes all normal and anomalous contractions and does not require a bond-operator representation of the spin correlator.
It is not, however, the exact correlator of the physical spin model with the local constraint imposed site by site.
Because the constraint is enforced only on average at the saddle-point level, the resulting correlator can violate the local-moment sum rule.

Equation~\eqref{eq:spin-correlator-bond-operator-identity} is an equal-time operator identity, whereas Eq.~\eqref{eq:bond-operator-spin-correlator} extends this decomposition to imaginary-time correlations through a factorization in terms of bond-operator two-point functions.
This defines the bond-operator correlation prescription used throughout the present work.
The upper representation in Eq.~\eqref{eq:Ising-representation-upper} is itself an exact operator identity and contains the $\mathcal A_{ij}$, $\mathcal B_{ij}$, $\mathcal C_{ij}^{\gamma}$, and $\mathcal D_{ij}^{\gamma}$ channels.
In the present prescription, choosing $\mathcal D_{ij}^{\gamma}$ as the finite triplet mean field fixes the bond-channel organization used to reconstruct the spin correlation: on a $\gamma$ bond, the bond-selected triplet mean-field channel is $\mathcal D_{ij}^{\gamma}$, instead of the $\mathcal D_{ij}^{\mu}$ channels with $\mu\neq\gamma$ that appear in the lower representation in Eq.~\eqref{eq:Ising-representation-lower}.
This organization is consistent with the bond-dependent structure adopted in the mean-field Ansatz and yields the AFM/FM momentum-space sign structure discussed in Ref.~\cite{Sasamoto-Nasu-2025}.
For the spin-averaged structure factor, the exact zeroth moment is
\begin{align}
    \label{eq:exact-spin-sum-rule}
    \frac{M}{N}\sum_{\bm q}^{\mathrm{B.Z.}}S(\bm q)
    =
    \frac{S(S+1)}{3}.
\end{align}
At $S=1$, both the Wick and bond-operator prescriptions yield unity for the integrated spectral weight, rather than the exact value $2/3$.
In the bond-operator prescription, the corresponding modified sum rule takes the form~\cite{Sasamoto-Nasu-2025}
\begin{align}
    \label{eq:bond-spin-sum-rule}
    \frac{M}{N}\sum_{\bm q}^{\mathrm{B.Z.}}S_{\mathrm{bond}}(\bm q)
    =
    \frac{S(S+1)}{2}.
\end{align}
Thus, for the present $S=1$ saddle, the zeroth-moment mismatch is shared by both evaluations and does not distinguish between the two correlation prescriptions.
It reflects a limitation of evaluating physical-spin correlations within the approximate mean-field treatment with the local constraint enforced only on average.
One may address the mismatch either by changing the spin length used to determine the saddle or by rescaling the response at fixed physical spin, but these procedures are inequivalent because the former changes the saddle itself~\cite{Mondal-2017,Mondal-2019,Rossi-Motruk-2023}.
In the present work, we retain the physical value $S=1$ in the self-consistent calculation and do not impose a common absolute-intensity rescaling between the SBMFT and ED spectra.

\section{Spinon dispersion and static structure factor in the pure Kitaev limit}
\label{app:a}

In this Appendix, we provide a complementary view of the difference between the $0$-flux and $\pi/2$-flux Ans\"atze in the pure $S=1$ Kitaev limit.
The purpose is to isolate two pieces of information that underlie the main-text dynamical results: the positive bosonic BdG eigenvalues of the mean-field Hamiltonian and the equal-time spin correlations encoded in the static structure factor.
Throughout this Appendix we set $\Gamma=\Gamma^{\prime}=0$ and use the self-consistent pure-Kitaev mean-field parameters.
At this point the hopping amplitudes and the off-diagonal triplet-pairing channels vanish within numerical precision, and the finite mean fields are the Lagrange multiplier and the bond-selected triplet amplitudes $\langle\mathcal{D}_{ij}^{\gamma}\rangle$ on $\gamma$ bonds.
For the two pure-Kitaev Ans\"atze considered here, let $D=|\langle\mathcal D_{ij}^{\gamma}\rangle|>0$ denote the common dimensionless magnitude on every $\gamma$ bond ($\gamma=x,y,z$). With the bond orientations in Fig.~\ref{fig:mean_field_ansatz}(a), $\langle\mathcal D_{ij}^{\gamma}\rangle=D$ for the $0$-flux Ansatz. For the $\pi/2$-flux Ansatz, it is $iD$ on the dotted bonds and $D$ on the remaining bonds. Reordering the Nambu components separates the $32\times32$ BdG matrix into four isospectral $8\times8$ blocks. In each block, the normal part is $\lambda\bm 1_4$. After removing an overall phase that does not change the singular values, the pairing part is $(KD/4)P_{\phi_t}(\bm k)$. With $u=e^{i\bm k\cdot(\bm\delta_x-\bm\delta_z)}$ and $v=e^{i\bm k\cdot(\bm\delta_y-\bm\delta_z)}$, where $\bm\delta_\gamma$ are nearest-neighbor bond vectors, a representative pair of dimensionless matrices is
\begin{align}
    P_0&=\begin{pmatrix}
        -u&iv&1&0\\
        iv&-u&0&1\\
        1&0&u&iv\\
        0&1&iv&u
    \end{pmatrix},
    \label{eq:zero-flux-pairing-block}\\
    P_{\pi/2}&=\begin{pmatrix}
        -u&iv&i&0\\
        iv&-iu&0&1\\
        1&0&u&-v\\
        0&1&iv&u
    \end{pmatrix}.
    \label{eq:pi2-flux-pairing-block}
\end{align}
Squaring the bosonic BdG dynamical matrix gives the positive spinon energies as
\begin{align}
    \varepsilon_{\bm k,j}^{2}
    =\lambda^2-\frac{(KD)^2}{16}\mu_{\bm k,j},
    \label{eq:pure-Kitaev-band-eigenvalues}
\end{align}
where $\mu_{\bm k,j}$ are the eigenvalues of $P_{\phi_t}(\bm k)P_{\phi_t}^{\dagger}(\bm k)$, namely the squared singular values of $P_{\phi_t}(\bm k)$, and $j=1,\ldots,4$ labels the positive-energy branches in one block.

The difference between the two flux patterns enters through these pairing singular values. Maximizing $\mu_{\bm k,j}$ over continuous momentum and branches gives $6$ for the $0$-flux Ansatz and $5+\sqrt3$ for the $\pi/2$-flux Ansatz. The spinon gaps therefore satisfy
\begin{align}
    \Delta_0^2&=\lambda^2-\frac{3}{8}(KD)^2,
    \label{eq:zero-flux-analytic-gap}\\
    \Delta_{\pi/2}^2&=\lambda^2-\frac{5+\sqrt3}{16}(KD)^2.
    \label{eq:pi2-flux-analytic-gap}
\end{align}
Each expression uses the $\lambda$ and $D$ of the corresponding self-consistent saddle. The gap closes at $\lambda/(|K|D)=\sqrt6/4$ for $0$ flux and $\sqrt{5+\sqrt3}/4$ for $\pi/2$ flux.

The gap difference is determined jointly by these flux-dependent coefficients and the self-consistent values of $\lambda$ and $D$. The weakly dispersive low-energy $\pi/2$-flux bands contribute more strongly to the boson occupation near condensation, allowing the $S=1$ constraint to be satisfied farther from gap closure. The parameters in Table~\ref{tab:pure-Kitaev-mean-fields} give $\Delta_0/|K|\simeq0.0126$ and $\Delta_{\pi/2}/|K|\simeq0.0679$, placing the $0$-flux saddle much closer to condensation. At finite $\Gamma$ and $\Gamma^{\prime}$, additional pairing fields enter, and the instability locations are determined by the self-consistent continuation described in Sec.~\ref{sec:Gamma-Gammap-dynamics}, not by these two-parameter formulas alone.

\begin{table}[t]
  \caption{Pure-Kitaev mean-field parameters and spinon gaps at $S=1$. The tabulated $D$ is the mean of the three bond-selected amplitudes. The gaps are minima over continuous momentum.}
  \label{tab:pure-Kitaev-mean-fields}
  \centering
  {
  \begin{tabular*}{\columnwidth}{@{\extracolsep{\fill}}lccc}
    \hline\hline
    Ansatz & $\lambda/|K|$ & $D$ & $\Delta/|K|$\\
    \hline
    $0$-flux & $0.596252$ & $0.973459$ & $0.0126$\\
    $\pi/2$-flux & $0.660761$ & $1.013270$ & $0.0679$\\
    \hline\hline
  \end{tabular*}}
\end{table}

\begin{figure*}[t]
  \centering
  \includegraphics[width=2\columnwidth]{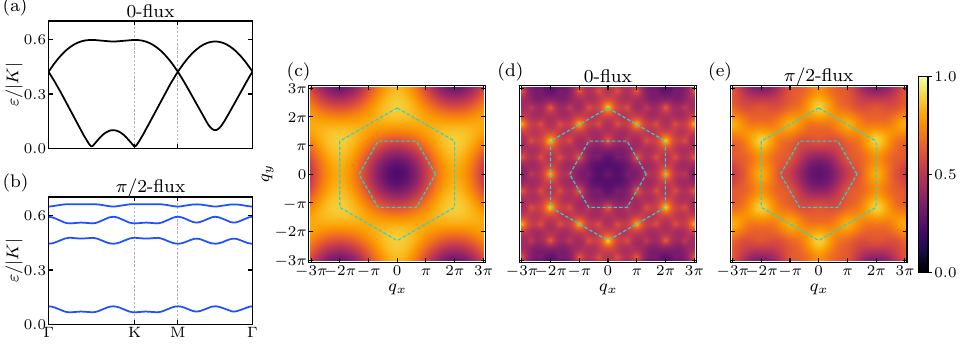}
  \caption{
  Spinon dispersions and static spin structure factors in the pure antiferromagnetic $S=1$ Kitaev limit.
  (a),(b) Positive bosonic BdG eigenvalues along the $\Gamma$--$\mathrm{K}$--$\mathrm{M}$--$\Gamma$ path for the $0$-flux and $\pi/2$-flux Ans\"atze, respectively.
  The calculation uses $\Gamma=\Gamma^{\prime}=0$ and keeps only the finite pure-Kitaev mean fields, namely the Lagrange multiplier and the bond-selected triplet pairings $\langle\mathcal{D}_{ij}^{\gamma}\rangle$.
  (c)--(e) Static spin structure factor $S(\bm{q})=[S^{xx}(\bm{q})+S^{yy}(\bm{q})+S^{zz}(\bm{q})]/3$.
  Panel (c) is the short-range reference obtained by retaining the on-site term and the nearest-neighbor Kitaev-bond correlation $\langle S_i^\gamma S_j^\gamma\rangle_{\gamma{\rm -bond}}\simeq -0.4330$, estimated from exact diagonalization on an $N=24$ cluster.
  Panels (d) and (e) show the corresponding SBMFT results for the $0$-flux and $\pi/2$-flux Ans\"atze.
  The same color scale is used in panels (c)--(e); panels (d) and (e) are normalized so that their maximum at the $\Gamma^{\prime}$ point coincides with that of panel (c).
  }
  \label{fig:spinon_disp_SSF}
\end{figure*}

Figures~\ref{fig:spinon_disp_SSF}(a) and \ref{fig:spinon_disp_SSF}(b) show that the two Ans\"atze already differ strongly at the level of the mean-field spinon dispersion.
For the $0$-flux Ansatz, the positive bosonic bands are highly dispersive and the lowest band has only a very small gap at $S=1$.
This is consistent with previous SBMFT studies, which found that the $0$-flux spin-liquid solution survives up to the vicinity of the physical spin-$1$ case but loses its gap and becomes unstable toward magnetic order at a slightly larger spin length, around $S\simeq1.07$~\cite{Ralko-Merino-2024,Sasamoto-Nasu-2025}.
By contrast, the $\pi/2$-flux Ansatz exhibits less dispersive positive bands and a visibly larger gap in the pure Kitaev limit.
Previous work also reported that this chiral bosonic solution remains gapped over a wider range of spin length, up to approximately $S\simeq2$ within SBMFT~\cite{Ralko-Merino-2024}.

The spinon dispersion itself is gauge dependent and is not directly observable.
Nevertheless, within SBMFT the dynamical spin structure factor is built from two-spinon processes generated by the diagonalized mean-field Hamiltonian, so the shape of the spinon bands strongly affects the organization of spectral weight in $S(\bm{q},\omega)$.
This explains the main-text contrast in Fig.~\ref{fig:DSSF_AFM_SBMFT}: the $0$-flux spectrum inherits the dispersive, tiny-gap character of the spinon bands, whereas the $\pi/2$-flux spectrum retains a gapped and weakly dispersive, Kitaev-like response.
In this restricted mean-field sense, the $\pi/2$-flux Ansatz is better suited to reproduce the flat dynamical structure characteristic of short-ranged Kitaev spin correlations.

The same tendency is also visible in the equal-time correlations.
The quantities shown in Figs.~\ref{fig:spinon_disp_SSF}(c)--\ref{fig:spinon_disp_SSF}(e) are the spin-averaged static structure factor $S(\bm q)$ defined in Eq.~\eqref{eq:trace spin structure factor}.
Figure~\ref{fig:spinon_disp_SSF}(c) shows a short-range reference constructed from the on-site contribution and the nearest-neighbor Kitaev-bond correlation obtained by exact diagonalization on an $N=24$ cluster~\cite{Koga-2018}.
It therefore represents the momentum-space profile expected when the real-space spin correlations are restricted to the short-range form characteristic of the pure Kitaev problem.
Figures~\ref{fig:spinon_disp_SSF}(d) and \ref{fig:spinon_disp_SSF}(e) show the static structure factors calculated within SBMFT for the $0$-flux and $\pi/2$-flux Ans\"atze, respectively.
For the comparison of the momentum dependence, each SBMFT result is multiplied by a single momentum-independent normalization factor so that its maximum value at the $\Gamma^{\prime}$ point coincides with that of the short-range reference in Fig.~\ref{fig:spinon_disp_SSF}(c).
This normalization places all three panels on the same color scale without changing the momentum profile of either SBMFT result.
The $\pi/2$-flux SBMFT result in Fig.~\ref{fig:spinon_disp_SSF}(e) closely follows this broad structure: the largest weight appears at the $\Gamma^{\prime}$ point and the intensity remains rather smoothly distributed over momentum space.
The $0$-flux result in Fig.~\ref{fig:spinon_disp_SSF}(d), in contrast, contains more visible oscillatory structure.
Thus, the static structure factor provides an equal-time counterpart of the dynamical comparison: within SBMFT, the $\pi/2$-flux Ansatz incorporates the short-range character of the Kitaev spin correlations more effectively than the $0$-flux Ansatz.
The smaller $0$-flux gap is consistent with its more extended mean-field correlations. The gap alone does not determine the physical-spin correlation range, which also depends on the spinon eigenvectors and the correlation prescription.

\section{Four-sublattice transformation and ferromagnetic spectra}
\label{app:four-sublattice-transformation}

\begin{figure}[t]
  \centering
  \includegraphics[width=\columnwidth]{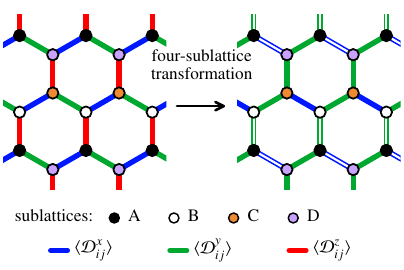}
  \caption{
  Schematic illustration of the four-sublattice transformation in the Schwinger-boson mean-field representation.
  In the left panel, the blue, green, and red bonds denote the AFM Kitaev mean fields $\langle\mathcal{D}_{ij}^{x}\rangle$, $\langle\mathcal{D}_{ij}^{y}\rangle$, and $\langle\mathcal{D}_{ij}^{z}\rangle$, respectively.
  The right panel shows the corresponding triplet-pairing components after the four-sublattice transformation.
  The transformed mean fields involve only the $\mathcal{D}^{x}$ and $\mathcal{D}^{y}$ components in the convention used here; open bonds indicate an additional minus sign.
  The black, white, orange, and purple sites denote the A, B, C, and D sublattices.
  }
  \label{fig:four_sub_transformation}
\end{figure}

\begin{figure*}[t]
  \centering
  \includegraphics[width=2\columnwidth]{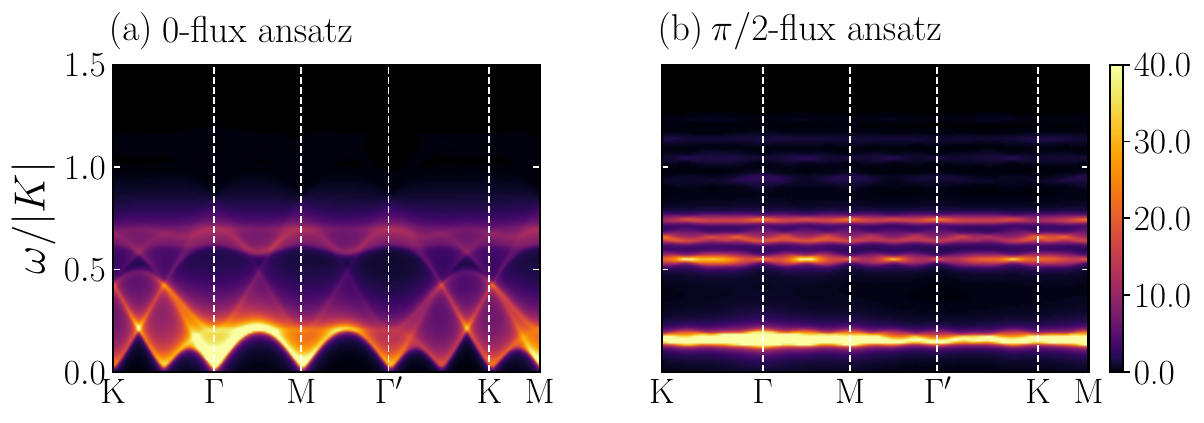}
  \caption{
  Zero-temperature dynamical spin structure factor $S(\bm{q},\omega)$ of the pure ferromagnetic $S=1$ Kitaev model calculated within SBMFT after implementing the four-sublattice mapping in the bond-operator convention.
  Panels (a) and (b) show the $0$-flux and $\pi/2$-flux Ans\"atze, respectively.
  The spectra are plotted along the same high-symmetry path as in Fig.~\ref{fig:DSSF_AFM_SBMFT}.
  The color scale, common to both panels, indicates the intensity of $S(\bm{q},\omega)$, and the broadening parameter is set to $\delta/|K|=0.01$.
  }
  \label{fig:DSSF_FM_SBMFT}
\end{figure*}

Here we summarize how the four-sublattice transformation introduced in Eq.~\eqref{eq:four-sublattice-unitary} acts in the Schwinger-boson representation.
We write the two-component Schwinger-boson spinor as
\begin{align}
    b_i=
    \begin{pmatrix}
        b_{i\uparrow}\\
        b_{i\downarrow}
    \end{pmatrix},
    \qquad
    S_i^\mu=\frac{1}{2}b_i^\dagger\sigma^\mu b_i .
\end{align}
For a site $i$ on sublattice $\Lambda=\mathrm{A},\mathrm{B},\mathrm{C},\mathrm{D}$, the spin rotation in Eq.~\eqref{eq:four-sublattice-unitary} can be represented on the spinor as
\begin{align}
    \widetilde{b}_i=u_{\Lambda}b_i ,
    \qquad
    u_{\mathrm{A}}=\sigma^0,\quad
    u_{\mathrm{B}}=i\sigma^x,\quad
    u_{\mathrm{C}}=i\sigma^y,\quad
    u_{\mathrm{D}}=i\sigma^z .
    \label{eq:appendix-four-sub-spinon-rotation}
\end{align}
With this gauge choice, the component form is
\begin{align}
\mathrm{A}:\quad&
\widetilde{b}_{i\uparrow}=b_{i\uparrow},\qquad
\widetilde{b}_{i\downarrow}=b_{i\downarrow},
\notag\\
\mathrm{B}:\quad&
\widetilde{b}_{i\uparrow}=i b_{i\downarrow},\qquad
\widetilde{b}_{i\downarrow}=i b_{i\uparrow},
\notag\\
\mathrm{C}:\quad&
\widetilde{b}_{i\uparrow}=b_{i\downarrow},\qquad
\widetilde{b}_{i\downarrow}=-b_{i\uparrow},
\notag\\
\mathrm{D}:\quad&
\widetilde{b}_{i\uparrow}=i b_{i\uparrow},\qquad
\widetilde{b}_{i\downarrow}=-i b_{i\downarrow}.
\label{eq:appendix-four-sub-spinon-components}
\end{align}
This representation gives $u_\Lambda^\dagger\sigma^\mu u_\Lambda=\eta_\Lambda^\mu\sigma^\mu$, with the signs $\eta_\Lambda^\mu$ defined below Eq.~\eqref{eq:spin transformation}.
It therefore reproduces the four-sublattice mapping between the AFM and FM pure Kitaev Hamiltonians discussed in Sec.~\ref{sec:Model}.
The overall sign of each $u_\Lambda$ is a local U(1) gauge choice for the Schwinger bosons and does not change the spin transformation.

We next record how the triplet-pairing mean fields used in the AFM Kitaev calculation are represented after this transformation.
This is useful because, in the AFM Ansatz, the finite mean field on a $\gamma$ bond is $\langle \mathcal{D}_{ij}^{\gamma}\rangle$.
After mapping the AFM state to the FM representation by the four-sublattice unitary transformation, the corresponding finite triplet-pairing component is obtained by transforming $\mathcal{D}_{ij}^{\gamma}$ with the spinor matrices in Eq.~\eqref{eq:appendix-four-sub-spinon-rotation}.
For the bond orientation and four-sublattice coloring used in Fig.~\ref{fig:honeycomb_lattice}(a), the nonzero AFM Kitaev mean fields are mapped as
\begin{align}
\mathrm{A}\mathrm{D}\text{-}x:\quad&
\mathcal{D}_{ij}^{x}\rightarrow \mathcal{D}_{ij}^{y},
&
\mathrm{C}\mathrm{B}\text{-}x:\quad&
\mathcal{D}_{ij}^{x}\rightarrow \mathcal{D}_{ij}^{y},
\notag\\
\mathrm{A}\mathrm{D}\text{-}y:\quad&
\mathcal{D}_{ij}^{y}\rightarrow -\mathcal{D}_{ij}^{x},
&
\mathrm{C}\mathrm{B}\text{-}y:\quad&
\mathcal{D}_{ij}^{y}\rightarrow \mathcal{D}_{ij}^{x},
\notag\\
\mathrm{A}\mathrm{B}\text{-}z:\quad&
\mathcal{D}_{ij}^{z}\rightarrow -\mathcal{D}_{ij}^{y},
&
\mathrm{C}\mathrm{D}\text{-}z:\quad&
\mathcal{D}_{ij}^{z}\rightarrow \mathcal{D}_{ij}^{y}.
\label{eq:appendix-four-sub-D-mapping}
\end{align}
This mapping is illustrated in Fig.~\ref{fig:four_sub_transformation}.
Only the triplet-pairing components relevant to the Kitaev mean fields are shown here.
The signs in Eq.~\eqref{eq:appendix-four-sub-D-mapping} are tied to the gauge choice in Eq.~\eqref{eq:appendix-four-sub-spinon-rotation}; changing the local U(1) gauge of the Schwinger bosons can move these signs among bonds without changing the spin transformation.
Equation~\eqref{eq:appendix-four-sub-D-mapping} shows that the transformed triplet-pairing component on a $\gamma$ bond is generally not $\mathcal{D}_{ij}^{\gamma}$ itself, but rather $\mathcal{D}_{ij}^{\mu}$ with $\mu\neq\gamma$.
Thus, although the AFM calculation in the main text uses the representation in Eq.~\eqref{eq:Ising-representation-upper}, where the energy gain on a $\gamma$ bond is naturally associated with $\mathcal{D}_{ij}^{\gamma}$, the four-sublattice-transformed FM representation is described by the alternative representation in Eq.~\eqref{eq:Ising-representation-lower}.
In that representation, a ferromagnetic Kitaev exchange gains energy from finite triplet-pairing mean fields in the components $\mu\neq\gamma$ on a $\gamma$ bond.
The two descriptions are nevertheless equivalent for the pure Kitaev model.
The four-sublattice transformation only permutes the triplet-pairing components and attaches gauge-dependent signs, leaving the magnitudes of the corresponding mean fields unchanged.
Therefore, when Eq.~\eqref{eq:Ising-representation-upper} is used for the AFM case and Eq.~\eqref{eq:Ising-representation-lower} for the FM representation, the SBMFT ground-state energies are mapped onto each other consistently with the unitary equivalence of the AFM and FM pure Kitaev Hamiltonians.
In this sense, the FM SBMFT is not obtained by merely changing the sign of $K$ while keeping the same $\mathcal{D}_{ij}^{\gamma}$ channel on a $\gamma$ bond.
Rather, one first maps the AFM bond fields by the four-sublattice transformation and then writes the FM Kitaev exchange in the representation of Eq.~\eqref{eq:Ising-representation-lower}, for which the finite triplet-pairing fields are the transformed components listed in Eq.~\eqref{eq:appendix-four-sub-D-mapping}.
The local Schwinger-boson constraint is unchanged by this transformation, and the self-consistent FM mean-field problem is therefore related to the AFM one by a relabeling of triplet components together with gauge-dependent bond signs.
Once the FM self-consistent solution is obtained in this convention, the spin correlations and the dynamical structure factor are evaluated by the same procedure as described in the main text.
For reference, the zero-temperature FM counterpart of the AFM full-SBMFT spectra in Fig.~\ref{fig:DSSF_AFM_SBMFT} is shown in Fig.~\ref{fig:DSSF_FM_SBMFT}.
The corresponding auxiliary short-range spectra for both AFM and FM cases are collected in Appendix~\ref{app:short-range-dynamics}.

\section{Short-range dynamical spectra}
\label{app:short-range-dynamics}

\begin{figure*}[t]
  \centering
  \includegraphics[width=2\columnwidth]{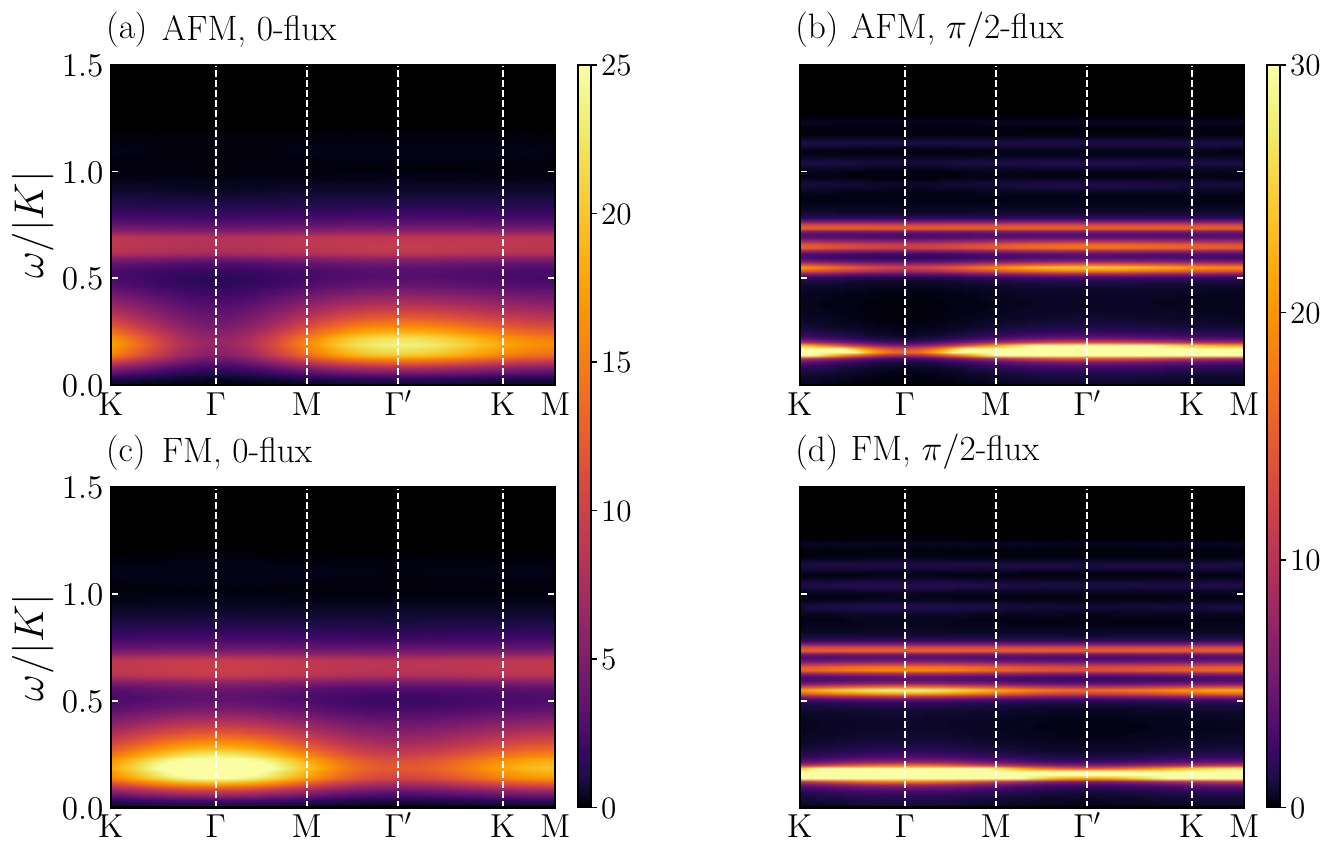}
  \caption{
  Auxiliary zero-temperature dynamical spin structure factors of the pure $S=1$ Kitaev model obtained after retaining only the on-site and nearest-neighbor spin correlations in the Fourier transform.
  Panels (a) and (b) show the antiferromagnetic Kitaev model for the $0$-flux and $\pi/2$-flux Ans\"atze, respectively.
  Panels (c) and (d) show the corresponding ferromagnetic spectra obtained in the four-sublattice-transformed bond-operator convention described in Appendix~\ref{app:four-sublattice-transformation}.
  The broadening parameter is set to $\delta/|K|=0.01$.
  }
  \label{fig:DSSF_short_T0_AF_FM}
\end{figure*}

In this Appendix, we collect the dynamical spin structure factors obtained by retaining only the on-site and nearest-neighbor spin correlations before performing the Fourier transform.
This calculation should be viewed as an auxiliary diagnostic of the short-range character of Kitaev spin correlations, rather than as a separate self-consistent mean-field Ansatz.
The same self-consistent SBMFT solutions are used as in the full spectra, but the real-space spin correlations entering $S(\bm{q},\omega)$ are restricted to the on-site and nearest-neighbor components.

Figures~\ref{fig:DSSF_short_T0_AF_FM}(a) and \ref{fig:DSSF_short_T0_AF_FM}(b) show the antiferromagnetic Kitaev case corresponding to the main-text full-SBMFT spectra in Fig.~\ref{fig:DSSF_AFM_SBMFT}.
For the $0$-flux Ansatz, the spectrum is dominated by nearly horizontal structures: a strong low-energy feature below $\omega/|K|\simeq0.3$ and a broader high-energy feature around $\omega/|K|\simeq0.6$--$0.7$.
Compared with the full SBMFT result in Fig.~\ref{fig:DSSF_AFM_SBMFT}(a), the fine momentum-dependent low-energy features are strongly suppressed, while the nearly horizontal low- and high-energy features become more prominent.
The $\pi/2$-flux result in Fig.~\ref{fig:DSSF_short_T0_AF_FM}(b) shows the same overall tendency.
The spectrum mainly consists of low- and high-energy features with weak momentum dependence; the low-energy feature is particularly flat, and the high-energy part appears as a broad feature around $\omega/|K|\simeq0.6$ with weak internal structure.
Thus, after restricting the Fourier transform to the short-range spin correlations, both AFM spectra are dominated by weakly momentum-dependent spectral weight, although the detailed high-energy distribution remains different between the two flux Ans\"atze.

Figures~\ref{fig:DSSF_short_T0_AF_FM}(c) and \ref{fig:DSSF_short_T0_AF_FM}(d) show the ferromagnetic counterparts obtained using the four-sublattice-transformed SBMFT convention summarized in Appendix~\ref{app:four-sublattice-transformation}.
The same short-range restriction again produces spectra dominated by weakly momentum-dependent low- and high-energy features.
The redistribution of intensity along the momentum path reflects the FM representation and the corresponding transformation of the bond operators, but the qualitative role of the short-range truncation is the same as in the AFM case.

\section{Zero-temperature exact-diagonalization calculation of the dynamical spin structure factor}
\label{app:Zero-temperature-exact-diagonalization-calculation-of-the-dynamical-spin-structure-factor}

In this Appendix, we briefly summarize how the zero-temperature dynamical spin structure factor $S(\bm{q},\omega)$ is calculated on a finite-size cluster by exact diagonalization, using the same Fourier convention as in Eq.~\eqref{eq:definition-dynamical-spin-structure-factor}.
The ED spectrum in Fig.~\ref{fig:ED_Sqw}(c) is given by
\begin{align}
\label{eq:ed_lehmann_sqw}
S(\bm{q},\omega)
&=\frac{1}{3}\sum_{\mu=x,y,z}S^{\mu\mu}(\bm{q},\omega)\notag\\
&=
\frac{2\pi}{3}\sum_{\mu=x,y,z}\sum_{n}
\left|
\langle n | S_{\bm{q}}^\mu |0\rangle
\right|^2
\delta\left(\omega-(E_n-E_0)\right),
\end{align}
where $|0\rangle$ and $|n\rangle$ are the ground state and excited eigenstates of Hamiltonian $\mathcal{H}$ with eigenvalues $E_0$ and $E_n$, respectively.
Here,
\begin{align}
\label{eq:ed_spin_operator_q}
S_{\bm{q}}^{\mu}
=
\frac{1}{\sqrt{N}}
\sum_{i}
e^{i\bm{q}\cdot\bm{r}_{i}}
S_{i}^{\mu}.
\end{align}
Instead of constructing all excited states explicitly, it is convenient to rewrite Eq.~\eqref{eq:ed_lehmann_sqw} in terms of the resolvent,
\begin{align}
\label{eq:ed_resolvent_sqw}
S^{\mu\mu}(\bm{q},\omega)
=
-2\,\mathrm{Im}
\left\langle 0 \left|
\left(S_{\bm{q}}^{\mu}\right)^\dagger
\frac{1}{\omega+i\eta+E_{0}-\mathcal{H}}
S_{\bm{q}}^{\mu}
\right|0\right\rangle.
\end{align}
For a finite cluster, the spectrum consists of discrete poles, and the finite $\eta$ replaces $2\pi\delta(\omega-\Delta E)$ by $2\eta/[(\omega-\Delta E)^2+\eta^2]$.
In the present calculations, we set $\eta/|K|=0.01$.
Introducing the normalized Krylov vector
\begin{align}
\label{eq:ed_lanczos_recursion}
|\phi_0\rangle
&=
\frac{S_{\bm{q}}^\mu|0\rangle}
{\sqrt{\langle 0|(S_{\bm{q}}^\mu)^\dagger S_{\bm{q}}^\mu|0\rangle}},
\notag\\
\mathcal{H}|\phi_{m}\rangle
&=
b_{m+1}|\phi_{m+1}\rangle
+a_m|\phi_m\rangle
+b_m|\phi_{m-1}\rangle,
\end{align}
with $a_{m}=\langle\phi_{m}|\mathcal{H}|\phi_{m}\rangle$ and $b_{0}=0$, one obtains a tridiagonal representation of $\mathcal{H}$ in the Krylov subspace generated from $S_{\bm{q}}^{\mu}|0\rangle$.
The corresponding resolvent matrix element is then written as the continued fraction
\begin{align}
\label{eq:ed_continued_fraction}
&\left\langle\phi_{0}\left|
\frac{1}{\omega+i\eta+E_{0}-\mathcal{H}}
\right|\phi_{0}\right\rangle\notag\\
&=
\cfrac{1}{z-a_{0}
-\cfrac{b_{1}^{2}}{z-a_{1}
-\cfrac{b_{2}^{2}}{z-a_{2}-\ddots}}},
\end{align}
where $z=\omega+i\eta+E_{0}$, and therefore
\begin{align}
S^{\mu\mu}(\bm{q},\omega)
&=
-2\,\mathrm{Im}\,
\Biggl[
\langle 0|(S_{\bm{q}}^\mu)^\dagger S_{\bm{q}}^\mu|0\rangle
\notag\\
&\quad\times
\left\langle \phi_0 \left|
\frac{1}{\omega+i\eta+E_{0}-\mathcal{H}}
\right|\phi_0\right\rangle
\Biggr].
\label{eq:ed_final_sqw}
\end{align}
This Krylov-space evaluation provides an efficient way to compute the zero-temperature $S(\bm{q},\omega)$ without explicitly constructing the full excitation spectrum.

For the pure Kitaev model, we use a periodic $N=24$ cluster with translations $(3,\sqrt{3})$ and $(0,2\sqrt{3})$ in the lattice units of Sec.~\ref{sec:Model}. We use the conserved plaquette fluxes to resolve the Hilbert space into symmetry sectors~\cite{Koga-2018}. The Lanczos calculation for the spin response is then restricted to the sectors reached by local spin operators acting on the ground state. We reconstruct the momentum-space spectrum from the on-site and bond-selected nearest-neighbor contributions~\cite{Baskaran-Sen-Shankar-2008}, using the lattice and spin symmetries of the cluster.

\bibliography{main}

\end{document}